\documentclass[preprint,nofootinbib,aps,superscriptaddress,eqsecnum]{revtex4-1}
\pdfoutput=1
\usepackage[colorlinks=true,citecolor=blue,linkcolor=blue,breaklinks=true]{hyperref}
\usepackage{graphicx}
\usepackage{mathrsfs}
\usepackage{soul}
\usepackage{natbib}
\usepackage[utf8]{inputenc}
\usepackage{enumerate}
\usepackage{amsmath,amssymb}
\usepackage{slashed}
\usepackage{amssymb,amsmath,subfigure,xcolor}

\newcommand{\lapprox}{%
\mathrel{%
\setbox0=\hbox{$<$}
\raise0.6ex\copy0\kern-\wd0
\lower0.65ex\hbox{$\sim$}
}}
\newcommand{\gapprox}{%
\mathrel{%
\setbox0=\hbox{$>$}
\raise0.6ex\copy0\kern-\wd0
\lower0.65ex\hbox{$\sim$}
}}

\newcommand{\ba}{\begin{array}}
\newcommand{\ea}{\end{array}}
\newcommand{\bd}{\begin{displaymath}}
\newcommand{\ed}{\end{displaymath}}
\newcommand{\beq}{\begin{equation}}
\newcommand{\eeq}{\end{equation}}
\newcommand{\bea}{\begin{eqnarray}}
\newcommand{\eea}{\end{eqnarray}}

\newcommand{\nn}{\nonumber}

\def\a{\alpha}

\def\b{\beta}

\def\m{\mu}

\def\q2 {q^2}

\def\bt{\begin{table}}
\def\et{\end{table}}

\catcode`@=11 
\def \gsim{\mathrel{\mathpalette\@versim>}}
\def \lsim{\mathrel{\mathpalette\@versim<}}
\def \@versim#1#2{\lower0.4ex\vbox{\baselineskip\z@skip\lineskip\z@skip
     \lineskiplimit\z@\ialign{$\m@th#1\hfil##\hfil$%
     \crcr#2\crcr\sim\crcr}}}
\catcode`@=12 

\begin{document}

\renewcommand*{\thefootnote}{\fnsymbol{footnote}}

\begin{center}

{\large\bf Unitarity constraints on 2HDM with  higher dimensional operators}\\[15mm] 
Deepak Sah\footnote{E-mail: \texttt{dsah129@gmail.com} \\ Present address: Raja Ramanna Centre for Advanced Technology, Indore-452013, India} 
\\[2mm]

{\em Discipline of Physics, Indian Institute of Technology Indore,\\
 Khandwa Road, Simrol, Indore - 453\,552, India}
\\[20mm]
\end{center}

\begin{abstract} 
\vskip 20pt

We study how the requirement of perturbative unitarity restricts the parameter space of the two-Higgs-doublet model (2HDM) when higher-dimensional operators up to dimension six are included. We demonstrate that such operators can enhance scalar production cross sections in vector boson fusion relative to 2HDM.
Using S-matrix unitarity, we place bounds on several dimension-six bosonic operators. We also find that certain “blind directions” in the Wilson coefficients of T-parameter-violating operators—which are poorly constrained by electroweak precision data—can be partially excluded when unitarity constraints are taken into account. These results demonstrate how high-energy consistency can complement experimental limits in defining the allowed parameter space of 2HDM effective field theory.


\end{abstract}
 \vskip 1 true cm
 \pacs {}
\maketitle

\section{Introduction}
\label{intro}

The Standard Model (SM) is in excellent agreement with a wide range of precision measurements from particle colliders such 
as the Large Electron–Positron Collider (LEP) and the Large Hadron Collider (LHC). However, it is widely recognized that the SM is incomplete, as it fails to address several fundamental theoretical and phenomenological puzzles, including the hierarchy and naturalness problems, the origin of neutrino masses, the nature of dark matter, and the observed baryon–antibaryon asymmetry of the universe. Many beyond-the-Standard-Model(BSM) frameworks attempt to address one or more of these issues by introducing additional fields, symmetries, or dynamical mechanisms. Among these, models that extend the scalar sector with a second Higgs doublet — collectively known as two-Higgs-doublet models (2HDMs) — appear naturally in a variety of well-motivated frameworks, such as supersymmetry, composite Higgs scenarios, and models with axions or extra dimensions.

The discovery of a Higgs boson with a mass near $125\ \text{GeV}$ and properties remarkably consistent with the SM expectations has sharpened the focus on 2HDMs as a minimal and phenomenologically rich benchmark for new-physics searches at colliders. In many realistic BSM scenarios, the second Higgs doublet is accompanied by additional heavy states. When these states are integrated out, their effects on the low-energy dynamics can be systematically described by higher-dimensional operators built from the 2HDM fields, leading to the framework of 2HDM effective field theory (2HDMEFT). A complete, non-redundant basis of dimension-six operators for 2HDMEFT has been constructed in the literature~\cite{Karmakar:2017yek,Crivellin:2016ihg,Anisha:2019nzx}, and phenomenological implications — for example, modifications of the alignment limit — have been explored in several studies~\cite{Karmakar:2018scg,Karmakar:2019vnq}.

A crucial theoretical constraint on any effective field theory is perturbative unitarity of the $S$-matrix. In the context of the SM effective field theory (SMEFT), unitarity considerations have been shown to place strong, model-independent bounds on the Wilson coefficients of higher-dimensional operators~\cite{Dahiya:2012ka}. Similar constraints are expected to apply in 2HDMEFT. When combined with experimental limits from Higgs coupling measurements, electroweak precision tests, and direct collider searches, unitarity bounds can significantly restrict the allowed parameter space of the model and help identify regions where the effective description remains self-consistent up to a given energy scale.

The standard tool for imposing perturbative unitarity on $2\to2$ scattering processes is the partial-wave expansion of the scattering amplitude,
\begin{equation}
\mathcal{M}(\theta) = 16\pi \sum_{\ell=0}^{\infty} (2\ell+1) \, a_{\ell} \, P_{\ell}(\cos\theta),
\label{eq:partialwave}
\end{equation}
where $\theta$ is the scattering angle and $P_{\ell}$ are the Legendre polynomials. From the orthonormality of the $P_{\ell}$, one can extract the partial-wave coefficients $a_{\ell}$ for a given process. For elastic scattering, the optical theorem implies the unitarity condition $|\text{Re}(a_{0})| < 1/2$ for the $\ell=0$ partial wave. This condition was famously applied by Lee, Quigg, and Thacker to derive an upper bound on the mass of the SM Higgs boson and to study the high-energy behavior of longitudinal vector-boson scattering~\cite{Lee:1977eg}. In the 2HDM, the same technique leads to powerful constraints on the quartic scalar couplings and, consequently, on the masses of the additional scalars~\cite{Akeroyd:2000wc}.

At energies well above the electroweak scale, the Goldstone boson equivalence theorem greatly simplifies the analysis of scattering amplitudes involving longitudinally polarized gauge bosons ($V_{L}$). It states that, to leading order in $M_V^2/s$, the amplitude for $V_L V_L \to V_L V_L$ scattering equals the amplitude for the corresponding scattering of the associated Goldstone bosons. In the CP-conserving 2HDM, the $S$-matrix for $2\to2$ bosonic processes factorizes into a block-diagonal form, comprising a $14\times14$ matrix for neutral channels and an $8\times8$ matrix for singly charged channels, giving a total of $22\times22$ independent two-body states. This structure imposes tight correlations among the quartic couplings of the scalar potential. When dimension-six operators are included, the block-diagonal pattern is generally preserved for $\varphi^4 D^2$ operators but can be disrupted by $\varphi^6$ operators, introducing new energy-dependent contributions that must be bounded by unitarity.

In this work, we systematically extend the unitarity analysis to the 2HDMEFT, focusing on the bosonic dimension-six operators of the types $\varphi^4 D^2$ and $\varphi^6$. We compute the full $S$-matrix for all $2\to2$ bosonic scattering channels, derive the resulting constraints on the Wilson coefficients and the new-physics scale $f$, and examine the interplay of these constraints with the experimentally favored alignment limit, $\cos(\beta-\alpha) \to 0$. In particular, we show that unitarity bounds are weakest near alignment, coinciding with the region preferred by LHC Higgs data. We also compare unitarity limits with existing experimental constraints from electroweak precision data (notably the $T$-parameter), Higgs signal strength measurements, and LHC searches for anomalous quartic gauge couplings. For custodial-symmetry-violating operators such as $O_{T1}$, $O_{T2}$, and $O_{T3}$, we identify and partially lift blind directions in the Wilson coefficient space by combining unitarity with $T$-parameter limits. Our results highlight the essential role of high-energy consistency in delineating the viable parameter space of 2HDMEFT and provide a set of constraints that complement current and future collider searches.


The paper is organized as follows. Sec. II introduces the 2HDMEFT framework. Sec. III computes \(V_L V_L \to V_L V_L\) amplitudes. Sec. IV derives unitarity constraints. Sec. V studies the alignment limit. Sec. VI compares with experimental bounds. Sec. VII concludes. Details are in the Appendix.

\section{Construction of the 2HDMEFT}
\label{sec:model}

\subsection{The 2HDM scalar sector}
The two scalar doublets are defined as 
\begin{equation}
\varphi_i = \frac{1}{\sqrt{2}} \begin{pmatrix} \sqrt{2} w_i^+ \\ (h_i+v_i) + i z_i \end{pmatrix}, \qquad i = 1,2 ,
\label{eqn1}
\end{equation}
where $w_i^{\pm}$, $h_i$, and $z_i$ denote the charged, neutral CP-even, and neutral CP-odd degrees of freedom, respectively, and $v_i$ is the vacuum expectation value (vev) of the $i$-th doublet.

Before spontaneous symmetry breaking (SSB), the tree-level 2HDM Lagrangian, extended by dimension-six operators, takes the form 
\begin{equation}
\mathcal{L} = \mathcal{L}_{\text{kin}} + \mathcal{L}_{\text{Yuk}} - V(\varphi_{1},\varphi_{2}) + \mathcal{L}_{6},
\label{eq:full_lag}
\end{equation}
with
\begin{align}
\mathcal{L}_{\text{kin}} &=
-\frac{1}{4} \sum_{X = G^{a},W^{i},B} X_{\mu\nu}X^{\mu\nu}
+ \sum_{I = 1,2} |D_{\mu} \varphi_{I}|^2
+ \sum_{\psi = Q,L,u,d,l} \bar{\psi} i \slashed{D} \psi, \\[4pt]
\mathcal{L}_{\text{Yuk}} &=
\sum_{I=1,2} Y^{e}_I \, \bar{l} \, e \,\varphi_{I}
+ \sum_{I=1,2} Y^{d}_I \, \bar{q} \, d \,\varphi_{I}
+ \sum_{I=1,2} Y^{u}_I \, \bar{q} \, u \,\tilde{\varphi}_{I}, \\[4pt]
V(\varphi_{1},\varphi_{2}) &=
m_{11}^2 |\varphi_{1}|^2 + m_{22}^2 |\varphi_{2}|^2
- \bigl( \mu^2 \varphi_{1}^{\dagger} \varphi_{2} + \text{h.c.} \bigr) \nonumber\\
&\quad + \lambda_1 |\varphi_{1}|^4 + \lambda_2 |\varphi_{2}|^4
+ \lambda_{3} |\varphi_{1}|^2 |\varphi_{2}|^2 \nonumber\\
&\quad + \lambda_4 |\varphi_{1}^{\dagger} \varphi_{2}|^2
+ \Bigl[ \Bigl( \frac{\lambda_5}{2} \varphi_{1}^{\dagger} \varphi_{2}
+ \lambda_6 |\varphi_{1}|^2 + \lambda_7 |\varphi_{2}|^2 \Bigr)
\varphi_{1}^{\dagger} \varphi_{2} + \text{h.c.}\Bigr], \\[4pt]
\mathcal{L}_{6} &= \sum_i \frac{c_i}{\,f^2}\, O_{i},
\label{eq:lag_terms}
\end{align}
where $c_i$ is the Wilson coefficient of the dimension-six operator $O_i$ and $f$ is the scale of new physics beyond the tree-level 2HDM. 
The terms proportional to $\lambda_{6,7}$ are referred to as ``hard-$Z_2$ violating'' because they induce a quadratically divergent amplitude for $\varphi_1 \leftrightarrow \varphi_2$ transitions~\cite{Ginzburg:2004vp} and can also introduce CP violation in the scalar sector~\cite{ElKaffas:2007rq}.
In this work, we restrict ourselves to the CP-conserving 2HDM and therefore set $\lambda_{6,7}=0$.
The vacuum expectation breaks electroweak symmetry values $v_1$ and $v_2$ of the two doublets $\varphi_{1,2}$.

In the CP-conserving case, the mass matrices of the neutral CP-even, neutral CP-odd, and charged scalars are diagonalized by the following field rotations:
\begin{equation}
\begin{pmatrix} H \\ h \end{pmatrix}
= R(\alpha) \begin{pmatrix} h_1 \\ h_2 \end{pmatrix}, \qquad
\begin{pmatrix} W_L^{\pm} \\ H^{\pm} \end{pmatrix}
= R(\beta) \begin{pmatrix} w_1^{\pm} \\ w_2^{\pm} \end{pmatrix}, \qquad
\begin{pmatrix} Z_L \\ A \end{pmatrix}
= R(\beta) \begin{pmatrix} z_1 \\ z_2 \end{pmatrix},
\label{eq:rotation}
\end{equation}
where
\begin{equation}
R(\theta) = \begin{bmatrix}
\cos \theta & \sin \theta \\
-\sin \theta & \cos \theta
\end{bmatrix}.
\end{equation}
The fields $h$ and $H$ are the physical neutral CP-even scalars, while $A$ and $H^{\pm}$ are the physical neutral CP-odd and charged scalars, respectively. As seen from Eq.~(\ref{eq:rotation}), $\beta$ is the mixing angle of the charged and CP-odd sectors, given by $\beta = \tan^{-1}(v_2/v_1)$. The mixing angle $\alpha$ of the CP-even neutral scalars can be expressed in terms of the elements of the CP-even mass-squared matrix $\mathcal{M}_{\rho}^2$, which in the gauge basis $(h_1,h_2)$ reads
\begin{equation}
\mathcal{M}_{\rho}^2 = \begin{pmatrix}
\mathcal{M}_{\rho 11}^2 & \mathcal{M}_{\rho 12}^2 \\[2pt]
\mathcal{M}_{\rho 12}^2 & \mathcal{M}_{\rho 22}^2
\end{pmatrix},
\end{equation}
with explicit expressions given in terms of the quartic couplings and vevs in the Appendix. The angle $\alpha$ is then obtained as
\begin{equation}
\alpha = \sin^{-1} \!\Biggl[ \,
\frac{\mathcal{M}_{\rho 12}^2}
{\sqrt{(\mathcal{M}_{\rho 12}^2)^2 + (\mathcal{M}_{\rho 11}^2 - m_{h}^2)^2}}
\,\Biggr].
\label{eq:alpha_expression}
\end{equation}

\subsection{Operators in 2HDMEFT}
\label{subsec:operators}

We adopt a complete basis of dimension-six operators for 2HDMEFT inspired by the SILH (Strongly Interacting Light Higgs) basis of SMEFT~\cite{Karmakar:2017yek}. The total Lagrangian can be written as
\begin{equation}
\mathcal{L} = \mathcal{L}_{2HDM}^{(4)}  + \mathcal{L}^{(6)},
\end{equation}
where the dimension-six part is organized by field content as
\begin{eqnarray}
\mathcal{L}^{(6)} &=& \mathcal{L}_{\varphi^4 D^2} + \mathcal{L}_{\varphi^2 D^2 X} + \mathcal{L}_{\varphi^2 X^2} + \mathcal{L}_{\varphi^6} + \mathcal{L}_{\varphi^3 \psi^2} + \mathcal{L}_{\varphi^2 \psi^2 D} + \mathcal{L}_{\varphi \psi^2 X} + \mathcal{L}_{D^2 X^2} + \mathcal{L}_{\psi^4}.\nn
\end{eqnarray}

Here $\varphi$, $\psi$, and $X$ denote the two Higgs doublets, fermion fields, and gauge field strength tensors, respectively, while $D$ stands for a covariant derivative. We follow the convention $\mathcal{L} \supset \frac{c_i}{f^2}O_i$, where the Wilson coefficient $c_i$ is named according to the corresponding operator $O_i$.

In this work we focus exclusively on bosonic operators, which can be categorized as follows:

\begin{itemize}
\item \textbf{$\boldsymbol{\varphi^6}$ operators} contain only Higgs doublets and modify the scalar potential. They represent corrections to the renormalizable 2HDM potential. The complete set of these operators, together with the modified minimization conditions, is listed in Appendix~\ref{app:potential}~\cite{Karmakar:2017yek}.

\item \textbf{$\boldsymbol{\varphi^4D^2}$ operators} involve four Higgs doublets and two derivatives. They redefine the kinetic terms of the Higgs fields, alter Higgs–gauge boson interactions, and affect the $W^\pm$ and $Z$ masses. The explicit forms of these operators are given in Appendix~\ref{app:phi4D2}.

\item \textbf{$\boldsymbol{\varphi^2X^2}$ operators} consist of two Higgs doublets and two field-strength tensors (e.g., $(\varphi^\dagger\varphi) X_{\mu\nu}X^{\mu\nu}$).

\item \textbf{$\boldsymbol{\varphi^2D^2X}$ operators} contain two Higgs doublets, two derivatives, and one field-strength tensor. These contribute to precision electroweak observables and to SM-like Higgs phenomenology.
\end{itemize}

The Wilson coefficients of $\varphi^2X^2$ and $\varphi^2D^2X$ operators are already constrained to be $\mathcal{O}(10^{-3})$ by measurements of the Higgs decay widths $h\to\gamma\gamma$ and $h\to Z\gamma$~\cite{Pomarol:2013zra}. 

Since such tight bounds render these operators phenomenologically less relevant for the present study, we focus on the classes $\varphi^4D^2$ and $\varphi^6$, which can induce sizable modifications in vector-boson scattering amplitudes while remaining less constrained by low-energy data.

\subsection{$V_L V_L \to V_L V_L$ scattering in 2HDMEFT}
\label{subsec:VLVL_scattering}

The scattering amplitude for $V_L V_L \rightarrow V_L V_L$ can be expanded in powers of the center-of-mass energy $E_{\text{cm}}$ as
\begin{equation}
\mathcal{M} = A_4 E_{\text{cm}}^4 + A_2 E_{\text{cm}}^2 + A_0 + A_{-2} E_{\text{cm}}^{-2} + \cdots ,
\end{equation}
where $E_{\text{cm}}$ is the center-of-mass energy. For a unitary theory, the coefficient $A_2$ of the quadratically growing term must vanish at energies $E_{\text{cm}} \gg M_i,(i = W, Z, h, H)$. This cancellation is achieved through the exchange of scalar particles in the model. When $A_2$ becomes zero, the theory is unitarized and the cross section decreases with energy. The gauge and scalar contributions to $A_2$ and $A_0$ are denoted as
\begin{equation}
A_2 = A_{2,g} + \sum_{S} A_{2,S},
\end{equation}
where $S = h, H$.

\subsubsection{Scattering amplitudes}
\label{subsubsec:scattering_amplitudes}

Following Ref.~\cite{Khan:2017xyh}, the expressions for the quadratically growing coefficient $A_2$ in the tree-level 2HDM are given below. In natural units $(\hbar = c = 1)$, $A_2$ has mass dimension $-2$ and is expressed in units of $[\mathrm{GeV}^{-2}]$.

The explicit forms for the leading $V_L V_L \rightarrow V_L V_L$ processes are:
\newline
1. $W_L^+ W_L^- \rightarrow W_L^+ W_L^-$
\begin{equation}
      A_2=  \frac{g_2^2(4 M_W^2-3 c_W^2 M_Z^2)(1+x)}{2  M_W^4}  -\frac{g_2^2}{2 M_W^2}~C^2~(1+x)
 \end{equation}
 2. $W_L^+ W_L^+ \rightarrow W_L^+ W_L^+$
 \begin{equation}
    A_2 =  \frac{g_2^2(3 c_W^2 M_Z^2-4 M_W^2)}{ M_W^4} + \frac{g_2^2}{M_W^2}~C^2 
 \end{equation}
 3. $W_L^+ W_L^- \rightarrow Z_L Z_L$
 \begin{equation}
   A_2 =  \frac{g_2^2 c_W^2 M_Z^2}{ M_W^4}  -\frac{g_2^2}{c_W M_W M_Z}~C C^{\prime} 
 \end{equation}
 
where $g_2$  is the $SU(2)_L$ gauge coupling, $M_V$ is the mass of the gauge boson $V = W^\pm, Z$, $c_W \equiv \cos\theta_W$, and $x \equiv \cos\theta$ with $\theta$ being the scattering angle, and the coupling multipliers are $C = \cos(\beta-\alpha)$ and $C' = \sin(\beta-\alpha)$.

The $\varphi^4 D^2$ operators induce a field redefinition of the Higgs fields~\cite{Karmakar:2018scg}, leading to a rescaling of the $hVV$ couplings:
 \begin{equation}
     \cos(\beta-\alpha) \rightarrow \cos(\beta-\alpha)(1-x_{2})+ \sin(\beta-\alpha) y
      \label{eq:cos_rescaling}
      \end{equation}
 \begin{equation}
    \sin(\beta-\alpha) \rightarrow  \sin(\beta-\alpha)(1-x_{1})+ \cos(\beta-\alpha) y
    \label{eq:sin_rescaling}
 \end{equation}
where $x_1$, $x_2$, and $y$ are functions of the Wilson coefficients of the higher-dimensional operators, given by
 \begin{eqnarray*}
 x_1 &=&\frac{v^2}{f^2} \Big( c_{H1} c_{\b}^2  s_{\a}^2 + c_{H2} c_{\a}^2 s_{\b}^2 + \frac{1}{8} c_{H1H2} s_{2\a} s_{2\b} + c_{H12} (c_{\a}^2 c_{\b}^2 + s_{\a}^2 s_{\b}^2 - \frac{1}{4} s_{2\a} s_{2\b})\\
&&\hspace{50pt}+ c_{H1H12} c_{\b} s_{\a} (s_{\a} s_{\b} - \frac{1}{2} c_{\a} c_{\b}) + c_{H2H12} c_{\a} s_{\b} (c_{\a} c_{\b} - \frac{1}{2} s_{\a} s_{\b}) \Big) ,\\
x_2 &=&\frac{v^2}{f^2} \Big( c_{H1} c_{\b}^2 c_{\a}^2 + c_{H2} s_{\a}^2 s_{\b}^2 + \frac{1}{8} c_{H1H2} s_{2\a} s_{2\b} + c_{H12} (s_{\a}^2 c_{\b}^2 + c_{\a}^2 s_{\b}^2 - \frac{1}{4} s_{2\a} s_{2\b})\\
&&\hspace{50pt}+ c_{H1H12} c_{\b} c_{\a} (c_{\a} s_{\b} - \frac{1}{2} s_{\a} c_{\b}) + c_{H2H12} s_{\a} s_{\b} (s_{\a} c_{\b} - \frac{1}{2} c_{\a} s_{\b}) \Big),\\
y &=& \frac{v^2}{f^2} \Big( \frac{1}{2} c_{H1} s_{2\a} c_{\b}^2 -\frac{1}{2} c_{H2}  s_{2\a} s_{\b}^2 - \frac{1}{8} c_{H1H2} c_{2\a} s_{2\b} - \frac{1}{2} c_{H12} (  c_{2\b}  s_{2\a} + \frac{1}{2} c_{2\a}  s_{2\b}) \\
&&\hspace{50pt}+ \frac{1}{4} c_{H1H12} ( s_{2\a} s_{2\b} - c_{2\a} c_{\b}^2 ) - \frac{1}{4} c_{H2H12} ( s_{2\a} s_{2\b} + c_{2\a} s_{\b}^2 )\Big).
\end{eqnarray*}
with $s_\theta \equiv \sin\theta$, $c_\theta \equiv \cos\theta$, etc. As a result of this field redefinition, the couplings of both CP-even neutral scalars to vector bosons and fermions are altered compared to the tree-level 2HDM. Using Eqs.~\eqref{eq:cos_rescaling} and \eqref{eq:sin_rescaling}, we obtain modified expressions for $A_2$ as functions of $\cos(\beta-\alpha)$ and $\tan\beta$ for various $V_L V_L$ scattering processes.


\newpage

\begin{figure}[ht]
\centering
\includegraphics[height=15em]{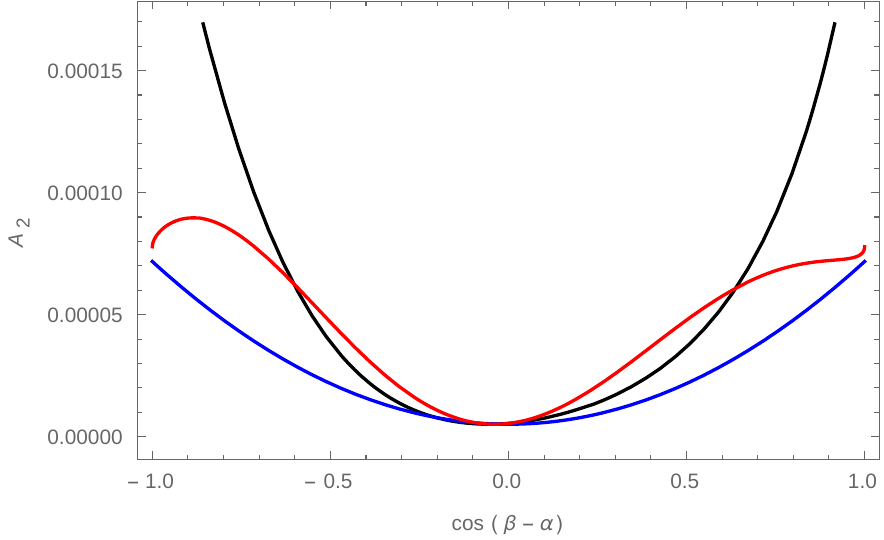}
\caption{The coefficient $A_2$ for $W_L^+ W_L^- \to W_L^+ W_L^-$ scattering as a function of $\cos(\beta-\alpha)$.
\textbf{Black  line:} Tree-level 2HDM (no dimension-six operators).
\textbf{Blue  line:} 2HDMEFT with $\varphi^4 D^2$ operators $O_{H1}$, $O_{H2}$, $O_{H12}$ ($c_{H1} = c_{H2} = c_{H12} = -1$, all other coefficients zero) and $\tan\beta = 1$.
\textbf{Red  line:} 2HDMEFT with same operators and coefficients as above but $\tan\beta = 5$.
Parameters: $\sqrt{s} = 2~\text{TeV}$, $f = 1~\text{TeV}$.}
\label{fig:A2_WW_WW}
\end{figure}

\begin{figure}[ht]
\centering
\includegraphics[height=15em]{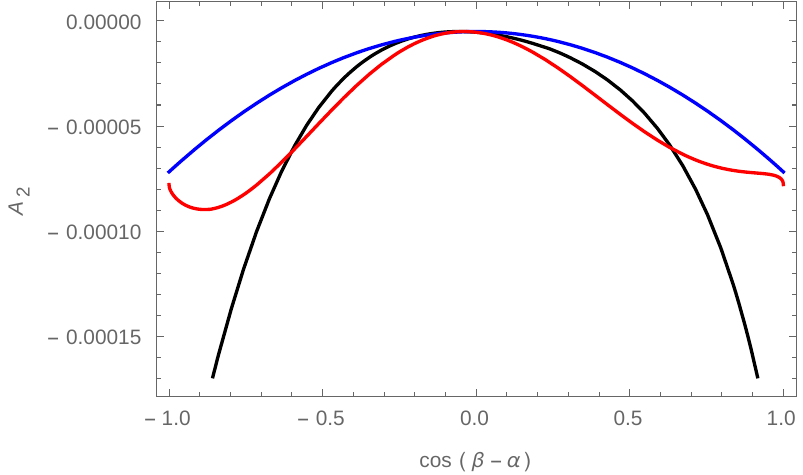}
\caption{The coefficient $A_2$ for $W_L^+ W_L^+ \to W_L^+ W_L^+$ scattering as a function of $\cos(\beta-\alpha)$.
\textbf{Black line:} Tree-level 2HDM.
\textbf{Blue line:} 2HDMEFT with $\varphi^4 D^2$ operators $O_{H1}$, $O_{H2}$, $O_{H12}$ ($c_{H1} = c_{H2} = c_{H12} = -1$, all other coefficients zero) and $\tan\beta = 1$.
\textbf{Red  line:} 2HDMEFT with same operators and coefficients as above but $\tan\beta = 5$.
Parameters: $\sqrt{s} = 2~\text{TeV}$, $f = 1~\text{TeV}$.}
\label{fig:A2_WWp_WWp}
\end{figure}

\begin{figure}[ht]
\centering
\includegraphics[height=15em]{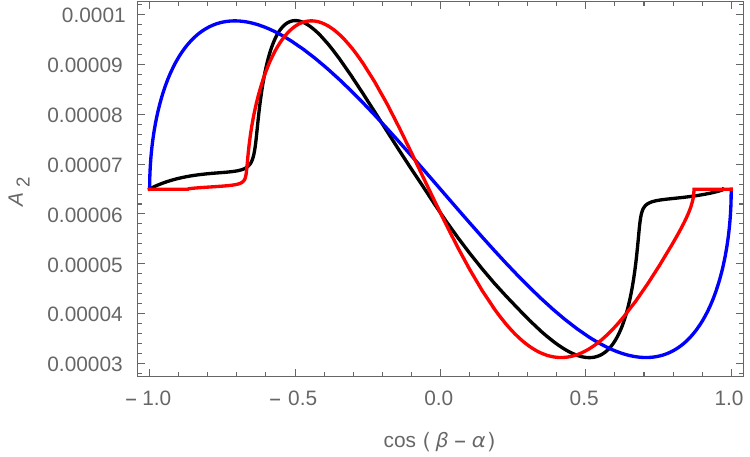}
\caption{The coefficient $A_2$ for $W_L^+ W_L^- \to Z_L Z_L$ scattering as a function of $\cos(\beta-\alpha)$.
\textbf{Black line:} Tree-level 2HDM.
\textbf{Blue  line:} 2HDMEFT with $\varphi^4 D^2$ operators $O_{H1}$, $O_{H2}$, $O_{H12}$ ($c_{H1} = c_{H2} = c_{H12} = -1$, all other coefficients zero) and $\tan\beta = 1$.
\textbf{Red line:} 2HDMEFT with same operators and coefficients as above but $\tan\beta = 5$.
Parameters: $\sqrt{s} = 2~\text{TeV}$, $f = 1~\text{TeV}$.}
\label{fig:A2_WW_ZZ}
\end{figure}

Figures~\ref{fig:A2_WW_WW}, \ref{fig:A2_WWp_WWp}, and \ref{fig:A2_WW_ZZ} illustrate the impact of dimension-six $\varphi^4 D^2$ operators on the high-energy growth coefficient $A_2$ (in units of $\text{GeV}^{-2}$) for vector boson scattering processes within the 2HDMEFT framework. In each figure, the tree-level 2HDM prediction (solid black line) is compared with 2HDMEFT results for two representative values of $\tan\beta$ (dashed blue: $\tan\beta=1$; dotted red: $\tan\beta=5$), with Wilson coefficients set to $c_{H1}=c_{H2}=c_{H12}=-1$.

The plots reveal a clear modification of $A_2$ relative to the tree-level case, particularly away from the alignment limit ($\cos(\beta-\alpha) \to 0$). This effect stems from the fact that $\varphi^4 D^2$ operators rescale the Higgs–gauge couplings through field redefinitions (Eqs.~17–18), thereby altering the roles of the scalar bosons $h$ and $H$ in unitarizing the high-energy amplitude.

The dependence on $\tan\beta$ is especially pronounced: larger values of $\tan\beta$ amplify deviations from the tree-level prediction, as evident in the red curves. This reflects the enhanced influence of the second Higgs doublet when $\tan\beta > 1$. Such modifications directly affect the corresponding scattering cross sections, which can be either enhanced or suppressed depending on the sign and magnitude of the Wilson coefficients. Consequently, experimental sensitivity to new scalar resonances in vector boson fusion channels at colliders such as the LHC may be significantly altered, making these figures essential for interpreting deviations from SM-like predictions in high-energy $VV$ scattering data.

The changes in $A_2$ due to dimension-six operators imply that the corresponding scattering cross sections are also modified in 2HDMEFT compared to the tree-level 2HDM. This can influence the discovery potential for new scalar resonances in high-energy collider searches.


\subsubsection{Cross sections}
\label{subsubsec:cross_sections}

\begin{figure}[ht]
\centering
\includegraphics[width=0.65\linewidth]{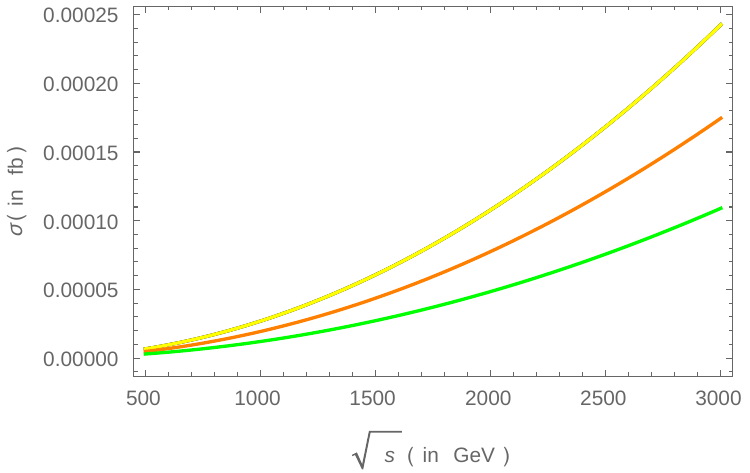}
\caption{Cross sections for vector-boson scattering processes at 
$\cos(\beta-\alpha)=0.5$, $f=1\;\text{TeV}$, $\tan\beta=5$, with 
$\varphi^{4}D^{2}$ operators $O_{H1}$, $O_{H2}$, and $O_{H12}$ active 
($c_{H1}=c_{H2}=c_{H12}=-1$, all other $\varphi^{4}D^{2}$ coefficients zero). 
\textbf{Yellow solid line:} $W_L^{\pm}W_L^{\mp}\to W_L^{\pm}W_L^{\mp}$; 
\textbf{Orange dashed line:} $W_L^{\pm}W_L^{\pm}\to W_L^{\pm}W_L^{\pm}$; 
\textbf{Green dotted line:} $W_L^{+}W_L^{-}\to Z_LZ_L$.}
\label{fig:cross_section_comparison}
\end{figure}

Figure~\ref{fig:cross_section_comparison} shows the cross sections as functions of the centre-of-mass energy $\sqrt{s}$ for three representative vector-boson scattering processes: 
$W_L^{\pm}W_L^{\mp}\to W_L^{\pm}W_L^{\mp}$ (yellow), 
$W_L^{\pm}W_L^{\pm}\to W_L^{\pm}W_L^{\pm}$ (orange), and 
$W_L^{+}W_L^{-}\to Z_LZ_L$ (green). 
The parameters are fixed as $\cos(\beta-\alpha)=0.5$, $\tan\beta=5$, and the new-physics scale $f=1\;\text{TeV}$. 
For illustration, we set the Wilson coefficients $C_{H1}=C_{H2}=C_{H12}=-1$, while all other coefficients are taken to be zero.

The curves exhibit the expected high-energy behaviour of massive gauge-boson scattering: a rise with $\sqrt{s}$ at moderate energies, followed by a turnover and a subsequent decrease at higher energies once the unitarising contributions from scalar exchange become dominant. 
The modifications induced by the $\varphi^{4}D^{2}$ operators are clearly visible: they shift the position of the maximum and alter the overall normalisation of the cross sections compared to the tree-level 2HDM (not shown). 
In particular, the chosen set of Wilson coefficients enhances the cross sections for the like-sign $W_L$ scattering ($W_L^{\pm}W_L^{\pm}\to W_L^{\pm}W_L^{\pm}$) and the $W_L^{+}W_L^{-}\to Z_LZ_L$ channel relative to the $W_L^{+}W_L^{-}\to W_L^{+}W_L^{-}$ process. 
This pattern illustrates how dimension-six operators can redistribute signal strengths among different vector-boson scattering channels, thereby affecting the experimental sensitivity to new scalars at high-energy colliders.

\section{Unitarity Constraints}
\label{sec:unitarity}

We consider all possible $2 \rightarrow 2$ bosonic elastic scattering processes. Any scattering amplitude can be expanded in partial waves as
\begin{equation}
\mathcal{M}(\theta) = 16\pi \sum_{\ell=0}^{\infty} a_\ell (2\ell+1) P_\ell(\cos\theta),
\end{equation}
where $\theta$ is the scattering angle and $P_\ell(x)$ is the Legendre polynomial of order $\ell$. 

The procedure is as follows: once the Feynman amplitude for a given $2\to 2$ process is computed, the partial-wave coefficients $a_\ell$ can be extracted using the orthonormality of the Legendre polynomials. This technique was first developed by Lee, Quigg, and Thacker for the SM~\cite{Lee:1977eg}, where they analyzed several two-body scatterings involving longitudinal gauge bosons and the physical Higgs boson.

The $\ell=0$ partial-wave amplitude $a_0$ is extracted from these amplitudes and arranged into an $S$-matrix whose rows and columns correspond to different two-body eigenstates. The largest eigenvalue of this matrix is constrained by the unitarity condition
\begin{equation}
| \operatorname{Re}(a_0) | < \frac12 .
\end{equation}
We now extend this method to the 2HDM with dimension-six operators (2HDMEFT). In this model, the same types of two-body scattering channels appear as in the tree-level 2HDM. We compute $a_0$ for every possible $2\to 2$ process and construct the corresponding $S$-matrix, taking the different two-body channels as rows and columns. First, we identify all possible two-particle channels, built from the fields $w_k^\pm$, $h_k$, and $z_k$ appearing in Eq.~\eqref{eqn1}. We consider neutral combinations (e.g., $w_i^+ w_j^-$, $h_i h_j$, $z_i z_j$, $h_i z_j$) and singly charged combinations (e.g., $w_i^+ h_j$, $w_i^+ z_j$).

The neutral-channel $S$-matrix for 2HDMEFT is a $14\times 14$ matrix with the following two-particle states as rows and columns:

\begin{eqnarray*}
|w_1^+w_1^->,~
|w_2^+w_2^->,~|w_1^+w_2^->,~|w_2^+w_1^->,~|\frac{h_1h_1}{\sqrt{2}}>,~|\frac{z_1z_1}{\sqrt{2}}>,
~|\frac{h_2h_2}{\sqrt{2}}>,~|\frac{z_2z_2}{\sqrt{2}}>,\end{eqnarray*}

\begin{eqnarray*}
~|h_1z_2>,~|h_2z_1>,~|z_1z_2>,~|h_1h_2>,~|h_1z_1>,~|h_2z_2>\,.
\end{eqnarray*}

The elements of this neutral-sector $S$-matrix are given in the appendix.

A similar construction applies to the singly charged two-particle states. The corresponding $S$-matrix is an $8\times 8$ matrix with the basis
\begin{eqnarray}
~|h_1 w_1^+>,~|h_1 w_2^+>,~|z_1 w_1^+>,
~|z_2 w_2^+>,\nn
\end{eqnarray}
\begin{eqnarray}
~|h_1 w_2^+>,~|h_2 w_1^+>,~|z_1 w_2^+>,
~|z_2 w_1^+>.\nn
\end{eqnarray}

The elements of the charged-sector $S$-matrix are also listed in the appendix.

Finding analytic expressions for the eigenvalues of these matrices is prohibitively difficult. We therefore solve the problem numerically. 
For coupled channels, the unitarity condition generalizez to requiring that the eigenvalues $\Lambda_i$ of the matrix of $a_0$ amplitudes satisfy $|\Lambda_i| \leq 1.$ In our normalization, this corresponds to the condition on the eigenvalues of the full scattering amplitude matrix $\mathcal{M}:$
\begin{equation}
|\Lambda_i| \le 8\pi .
\end{equation}
We have implemented the modifications due to the dimension-six operators into the public code 2HDMC~\cite{Eriksson:2009ws}, facilitating the verification of unitarity in the presence of these higher-dimensional terms.

\subsection{$\varphi^6$ operators}
\label{sec:scalar}
In the tree-level 2HDM, the $S$-matrix for $2\to 2$ bosonic scattering is a block-diagonal $22\times 22$ matrix, decomposing into submatrices for neutral channels (two $6\times6$ blocks and one $2\times2$ block) and for charged channels (two $4\times4$ blocks). This block structure is altered by the $\varphi^6$ operators. With these operators included, the $S$-matrix becomes a non-block-diagonal $22\times 22$ matrix. However, because its entries are proportional only to $\cos\beta$, $\sin\beta$ (which are bounded by $1$) and to $v_{1,2}$ (which are much smaller than the new-physics scale $f$), the resulting eigenvalues lead to unitarity bounds that are significantly weaker than those in the tree-level 2HDM case.

\subsection{$\varphi^4 D^2$ operators}
\label{subsec:phi4D2_unitarity}

In the presence of $\varphi^4 D^2$ operators, the $S$-matrix for $2\to 2$ bosonic scattering receives additional contributions, while the block-diagonal structure of the tree-level 2HDM $S$-matrix is preserved. These extra contributions, given in Appendix~\ref{amplitudes}, are proportional to $s/f^2$, where $s = (\sum p_{\text{in}})^2$ and $p_{\text{in}}$ denotes the four-momenta of the incoming particles.

\begin{figure}[h!]
\centering
\subfigure[]{\includegraphics[width=2.5in,height=2in]{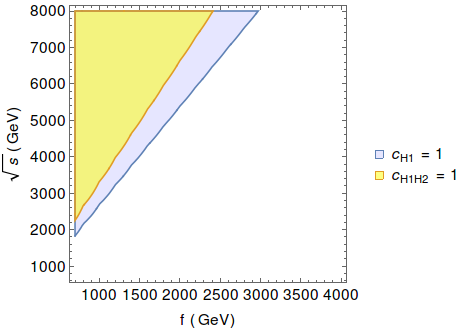}}
\hskip 10pt
\subfigure[]{\includegraphics[width=2.5in,height=2in]{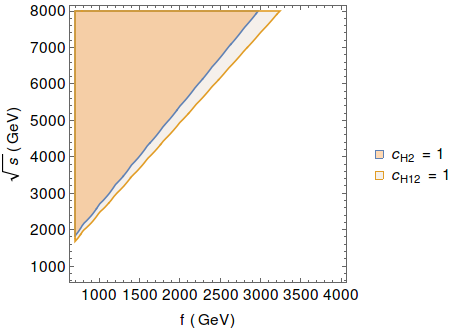}} \\
\vspace{10pt}
\subfigure[]{\includegraphics[width=2.5in,height=2in]{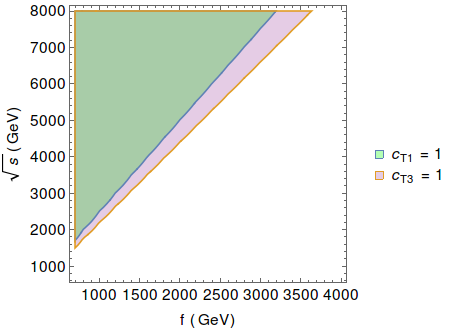}}
\hskip 10pt
\subfigure[]{\includegraphics[width=2.5in,height=2in]{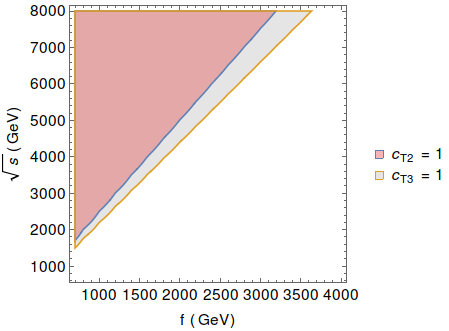}}
\caption{Unitarity bounds in the $\sqrt{s}$–$f$ plane for different Wilson coefficients of $\varphi^4 D^2$ operators. Blue shaded regions are excluded by perturbative unitarity ($|\Lambda_i| > 8\pi$); white regions are allowed.}
\label{fig:unitarity_plane}
\end{figure}

Applying the unitarity condition $|\Lambda_i| \le 8\pi$, we directly obtain bounds on $\sqrt{s}$. Figure~\ref{fig:unitarity_plane} shows the resulting constraints in the $\sqrt{s}$–$f$ plane for several representative Wilson coefficients; the blue regions are excluded by perturbative unitarity.

The complete set of $2\to 2$ scattering amplitudes, many of which are related through Wick's theorem, is provided in appendix ~\ref{amplitudes}. The unitarity condition $|\Lambda_i| \le 8\pi$ directly translates into upper limits on $\sqrt{s}$ for a given new-physics scale $f$.
The unitarity bounds in the $\sqrt{s}$--$f$ plane are shown for selected Wilson coefficients of the $\varphi^{4}D^{2}$ operators in the 2HDMEFT framework. The blue shaded regions are excluded by the condition $|\Lambda_{i}| > 8\pi$ applied to the eigenvalues of the zeroth partial-wave matrix for all $2\to 2$ bosonic scattering channels. Each curve corresponds to a distinct choice of operator coefficients, illustrating how different combinations of higher-dimensional terms modify the high-energy behavior of scattering amplitudes. As expected, for a fixed new-physics scale $f$, unitarity violation occurs at lower $\sqrt{s}$ when the Wilson coefficients are larger, reflecting the $s/f^{2}$ growth of the dimension-six contributions. Conversely, increasing $f$ suppresses these effects, extending the regime of validity of the effective description to higher center-of-mass energies.

These exclusion contours highlight the interplay between the cutoff scale of the effective theory and the allowed energy range for scattering processes. In particular, for typical values $f \sim 1\;\text{TeV}$, perturbative unitarity is generally lost at $\sqrt{s} \sim 2$--$3\;\text{TeV}$, a range accessible at the LHC in vector-boson fusion channels. The variation among the curves underscores that operators with different Lorentz or gauge structures can lead to markedly different unitarity limits, thereby providing a model-independent diagnostic of which effective interactions are most constrained by high-energy consistency. Such bounds are complementary to low-energy precision tests and direct collider searches, serving as an essential criterion for delineating the viable parameter space of the 2HDMEFT.

If all new physics (NP) fields are heavy, their effects on the 2HDM dynamics can be parametrized by a set of higher-dimensional operators. These operators spoil the renormalizability of the effective theory and can cause certain scattering amplitudes to violate unitarity. The resulting bounds scale as $f^2$ and also depend on $\sqrt{s}$, becoming stronger at higher energies. For $\varphi^4 D^2$-type operators, unitarity typically bounds $\sqrt{s}$ around 2~TeV, a typical parton-level energy at the LHC. Combining unitarity constraints with $T$-parameter measurements yields stringent bounds on the Wilson coefficients.

\subsection{Bounds on $T$-parameter violating operators}
\label{subsec:T_param_bounds}

Electroweak precision observables, most notably the $T$-parameter, place stringent constraints on operators that break custodial symmetry. In our basis, the operators $O_{T1}$, $O_{T2}$, and $O_{T3}$ contribute to the $T$-parameter at tree level. Combining the high‑energy constraints from perturbative unitarity with the low‑energy bounds from the $T$-parameter yields complementary limits on the corresponding Wilson coefficients.

\begin{figure}[h!]
\centering
\subfigure[$(C_{T1},C_{T3})$ plane]{\includegraphics[width=2.0in,height=2.0in]{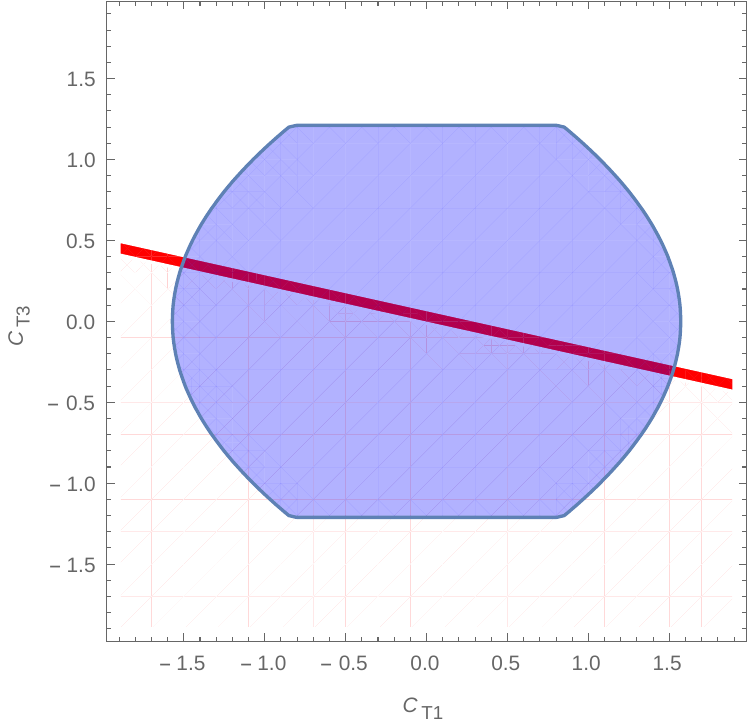}}
\hskip 15pt
\subfigure[$(C_{T2},C_{T3})$ plane]{\includegraphics[width=2.0in,height=2.0in]{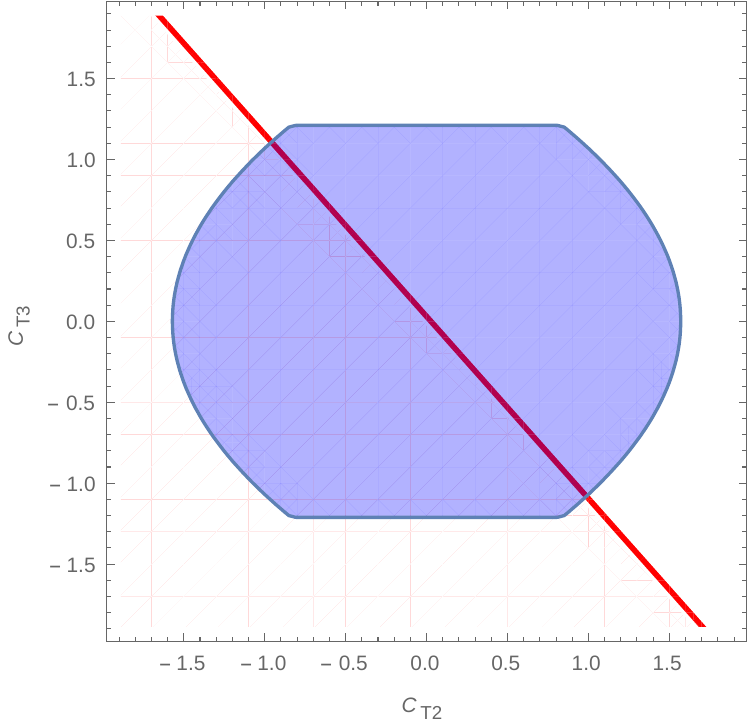}}
\caption{
Joint constraints on Wilson coefficients of custodial-symmetry-violating $\varphi^4 D^2$ operators. \\
\textbf{Panel (a):} $c_{T1}$ vs. $c_{T3}$ for operators $O_{T1}$ and $O_{T3}$. \\
\textbf{Panel (b):} $c_{T2}$ vs. $c_{T3}$ for operators $O_{T2}$ and $O_{T3}$. \\
\textbf{Blue shaded regions:} Allowed by unitarity ($|a_0|<1$ at $\sqrt{s}=2~\text{TeV}$). \\
\textbf{Red shaded regions:} Allowed by $T$-parameter constraints. \\
Parameters: $f=1~\text{TeV}$, $\tan\beta=5$, $\cos(\beta-\alpha)=0.1$.}
\label{fig:T_operator_bounds}
\end{figure}

Figure~\ref{fig:T_operator_bounds} displays the allowed regions in the $(C_{T1},C_{T3})$ plane (left panel) and the $(C_{T2},C_{T3})$ plane (right panel), obtained by imposing two independent conditions:
\begin{enumerate}
\item perturbative unitarity, $|a_0|<1$, applied to $2\to2$ bosonic scattering at $\sqrt{s}=2~\text{TeV}$ (blue regions), and
\item the experimental limit on the $T$-parameter from global electroweak fits (red regions).
\end{enumerate}
The new‑physics scale is fixed to $f=1~\text{TeV}$.

The blue regions represent the parameter space for which the effective theory remains unitary up to $\sqrt{s}=2~\text{TeV}$. The red bands indicate the values of the Wilson coefficients consistent with the measured $T$-parameter at 95\% CL. As seen in the figure, the unitarity‑allowed (blue) and $T$-parameter‑allowed (red) regions are largely disjoint for the chosen scale $f=1~\text{TeV}$; only a narrow intersection near the origin may be present. This indicates that, at this scale, the combination of unitarity and electroweak precision data strongly restricts the allowed range of the custodial‑violating Wilson coefficients, with the two constraints acting in complementary directions in parameter space.

The figure illustrates key features of the constraints: unitarity and the $T$-parameter probe different energy regimes and exclude largely orthogonal regions of the Wilson‑coefficient space. Their combination leaves only a very limited allowed region near the origin. While the $T$-parameter imposes strong one‑dimensional bounds on individual coefficients, unitarity further restricts correlated deviations. The small overlap (or absence thereof) highlights the powerful synergy between high‑energy consistency and low‑energy precision tests.

Thus, even for operators that are tightly constrained by electroweak precision data, unitarity provides essential, model‑independent restrictions that can eliminate remaining allowed directions and significantly tighten the viable parameter space of the 2HDMEFT.

 \section{The Alignment Limit and Its Implications for Unitarity}

The alignment limit, defined by \(\cos(\beta - \alpha) \to 0\), is of particular phenomenological importance in 2HDM scenarios, as it ensures that the couplings of the light CP-even scalar \(h\) to gauge bosons and fermions match their Standard Model (SM) values. This limit is strongly favored by LHC Higgs measurements, which constrain \(|\cos(\beta - \alpha)| \lesssim 0.1\) at 95\% CL for most 2HDM types ~\cite{Bernon:2015qea}\cite{Bernon:2015wef}.


In the tree-level 2HDM, the coupling multipliers for the CP-even neutral scalars to vector bosons are
\begin{equation}
    \kappa_{hVV} = \sin(\beta - \alpha), \qquad
\kappa_{HVV} = \cos(\beta - \alpha),
\end{equation}

where \(V = W, Z\). In the alignment limit (\(\cos(\beta - \alpha) = 0\), \(\sin(\beta - \alpha) = 1\)), the light Higgs \(h\) couples to gauge bosons exactly as in the SM (\(\kappa_{hVV} = 1\)), while the heavy scalar \(H\) decouples (\(\kappa_{HVV} = 0\)). This decoupling has profound consequences for the unitarization of vector boson scattering.

When dimension-six operators of type \(\varphi^{4}D^{2}\) are included, these couplings are modified as

\[
\kappa'_{hVV} = (1 - x_{1})\sin(\beta - \alpha) + y\cos(\beta - \alpha),
\]
\[
\kappa'_{HVV} = (1 - x_{2})\cos(\beta - \alpha) + y\sin(\beta - \alpha),
\]

where \(x_{1}\), \(x_{2}\), and \(y\) are functions of the Wilson coefficients. At exact alignment, these reduce to

\[
\kappa'_{hVV} = 1 - x_{1}, \qquad
\kappa'_{HVV} = y .
\]

Thus, even at alignment, the dimension-six operators can alter the \(hVV\) coupling through \(x_{1}\), while \(H\) acquires a coupling to gauge bosons proportional to \(y\).


The high-energy behavior of \(V_{L}V_{L} \to V_{L}V_{L}\) scattering amplitudes is especially sensitive to the alignment condition. Consider the dominant channel \(W^{+}_{L}W^{-}_{L} \to W^{+}_{L}W^{-}_{L}\), whose leading energy-growing term is given by

\[
A_{2} = \frac{g_{2}^{2}(4M_{W}^{2} - 3c_{W}^{2}M_{Z}^{2})(1+\cos\theta)}{2M_{W}^{4}} - \frac{g_{2}^{2}}{2M_{W}^{2}} C^{2}(1+\cos\theta),
\]

where \(C = \cos(\beta - \alpha)\).

- At the alignment limit (\(C = 0\)):
  The term proportional to \(C^{2}\) vanishes, and the heavy scalar \(H\) does not contribute to unitarization. The amplitude is unitarized solely through SM-like Higgs exchange (\(h\)). Any remaining high-energy growth arises only from the dimension-six operators themselves.

- Away from alignment (\(|C| > 0\)):
  The \(C^{2}\) term contributes to the high-energy growth, both \(h\) and \(H\) participate in unitarization, and interference between tree-level 2HDM contributions and dimension-six operators enhances the amplitude. Consequently, unitarity constraints become more stringent.

The zeroth partial wave amplitude can be expressed schematically as

\[
a_{0} \approx \frac{s}{16\pi}\left[-\frac{g_{2}^{2}}{2M_{W}^{2}}C_{\text{eff}}^{2} + \frac{c_{i}}{f^{2}}\right] + \text{constant terms},
\]

where \(C_{\text{eff}}\) is the effective coupling modified by dimension-six operators. The unitarity condition \(| \operatorname{Re}(a_{0}) | < 1/2\) then translates into bounds on the Wilson coefficients that depend strongly on \(\cos(\beta - \alpha)\).


To illustrate this dependence, we evaluate the maximum allowed values of three representative Wilson coefficients—\(c_{H1}\), \(c_{H2}\), and \(c_{H12}\)—as functions of \(\cos(\beta - \alpha)\), derived from the unitarity condition applied to \(W^{+}_{L}W^{-}_{L} \to W^{+}_{L}W^{-}_{L}\) scattering with \(\sqrt{s} = 2\ \text{TeV}\), \(f = 1\ \text{TeV}\), and \(\tan\beta = 5\).

- At exact alignment (\(\cos(\beta - \alpha) = 0\)):
  \[
  |c_{H1}|_{\text{max}} \approx 6.3,\quad
  |c_{H2}|_{\text{max}} \approx 1.2,\quad
  |c_{H12}|_{\text{max}} \approx 4.5 .
  \]

- At \(\cos(\beta - \alpha) = 0.1\) (experimentally favored region):
\begin{align*}
     |c_{H1}|_{\text{max}} \approx 1.8\ (\text{about}\  3.5\ \text{times tighter}),\\
  |c_{H2}|_{\text{max}} \approx 1.0\ (\text{about}\ 1.2\ \text{times tighter}),\\
  |c_{H12}|_{\text{max}} \approx 2.5\ (\text{about}\ 1.8\ \text{times tighter}) .
\end{align*}

The coefficient \(c_{H1}\), associated with the operator \(O_{H1} = (\partial_{\mu}|\varphi_{1}|^{2})^{2}\), shows the strongest alignment dependence because \(\varphi_{1}\) decouples from the SM-like Higgs at alignment, relaxing constraints. Away from alignment, mixing enhances its contribution to \(V_{L}V_{L}\) scattering, tightening the bounds. The different behaviors of \(c_{H2}\) and \(c_{H12}\) reflect their distinct roles in the scalar sector.


Current LHC measurements constrain \(|\cos(\beta - \alpha)| \lesssim 0.1\) ~\cite{Bernon:2015wef}. This low-energy preference has important implications for high-energy consistency:
 Relaxed unitarity bounds: The experimentally favored region near alignment coincides with the weakest unitarity constraints.
 \newline
Consistency window: For a given Wilson coefficient, the maximum \(\sqrt{s}\) up which the effective theory remains unitary is larger near alignment.
\newline
Complementarity:Higgs precision measurements (constraining \(\cos(\beta - \alpha)\)) and unitarity considerations (constraining \(c_{i}\)) provide complementary probes of the 2HDMEFT parameter space.

Thus, the alignment limit not only ensures SM-like Higgs couplings but also extends the regime of validity of the effective theory in the presence of higher-dimensional operators. This creates a theoretically consistent window for sizable new physics effects in the region most compatible with current experimental data.










\section{Comparison with Experimental Constraints}
\label{sec:experimental}
The unitarity bounds derived from the high-energy behavior of the 2HDMEFT must be contextualized alongside existing experimental limits from electroweak precision data (EWPD), Higgs signal strength measurements, and direct searches for anomalous quartic gauge couplings (aQGCs) in vector boson fusion (VBF) processes at the LHC. While EWPD and Higgs data constrain the low-energy parameter space of the effective theory, our unitarity analysis provides fundamental high-energy consistency conditions that are complementary and, in some cases, the leading constraints for operators poorly probed at low energies.

Electroweak precision constraints are particularly sensitive to operators that violate custodial symmetry or modify the $W$ and $Z$ self-energies. In our basis, the operators $O_{T1}$, $O_{T2}$, and $O_{T3}$ contribute to the $T$-parameter at tree level. The global fit to EWPD imposes a stringent bound of approximately $|c_T| \lesssim 10^{-3}$ for $f = 1\ \text{TeV}$ \cite{Baak:2014ora,deBlas:2019qco}, which is orders of magnitude stronger than the corresponding unitarity bounds shown in Fig.~6. However, many $\varphi^4 D^2$ operators (e.g., $O_{H1}$, $O_{H12}$) do not contribute to the $T$-parameter and are only weakly constrained by EWPD.

Higgs signal strength measurements at the LHC constrain deviations of the SM-like Higgs couplings to gauge bosons and fermions. In the 2HDMEFT, the $\varphi^4 D^2$ operators rescale the $hVV$ couplings as shown in Eqs.~(12)–(13). Current global fits limit deviations in the Higgs coupling scale factors to the level of $|\kappa_V - 1| \lesssim 0.05$–$0.10$ \cite{ATLAS:2019nkf,CMS:2021klu}, which translates to bounds on combinations of Wilson coefficients at the level of $|c_i| \lesssim 0.1$–$1$ for $f = 1\ \text{TeV}$. These bounds are typically stronger than unitarity limits for the same operators when $\cos(\beta-\alpha)$ is small (alignment limit), but unitarity becomes increasingly important as one moves away from alignment.

Direct searches for aQGCs in VBF processes such as $pp \to jj W^+ W^-$ or $jj ZZ$ at the LHC probe the same high-energy $V_L V_L \to V_L V_L$ scattering amplitudes that underlie our unitarity analysis \cite{ATLAS:2022zbu,CMS:2021jji}. The experimental limits are often expressed as bounds on the coefficients $f_{T,i}/\Lambda^4$ of the Warsaw-basis aQGC operators. These can be mapped to our $\varphi^4 D^2$ operators, yielding approximate bounds of $|c_i| \lesssim \mathcal{O}(10)$ for $f = 1\ \text{TeV}$ at 95\% CL. While these direct collider bounds are currently comparable to or weaker than our unitarity constraints (see Fig.~5), they provide independent, data-driven limits that will improve with future LHC data.
Scalar self-interactions governed by $\varphi^6$ operators are only weakly probed by current experiments through di-Higgs or triple-Higgs production, which have limited sensitivity. For these operators, unitarity often provides the most meaningful constraint on the allowed parameter space, as shown in Sec.~\ref{sec:unitarity}.

The following table summarizes the comparative strength of unitarity bounds against existing experimental constraints for representative operators in the 2HDMEFT.

\begin{table}[t]
\centering
\footnotesize
\caption{Comparison of unitarity bounds with existing experimental constraints for selected 2HDMEFT operators ($f = 1~\mathrm{TeV}$).}
\label{tab:constraints}

\begin{tabular*}{\textwidth}{@{\extracolsep{\fill}}
p{3.2cm} p{2.8cm} p{2.6cm} p{2.9cm} p{1.8cm}}
\hline\hline
\textbf{Operator} &
\textbf{Probe} &
\textbf{Exp. bound} &
\textbf{Unitarity bound} &
\textbf{Dominant} \\
& & & ($\sqrt{s}=2~\mathrm{TeV}$) & \\
\hline
$\varphi^4 D^2$ (Custodial violation,
e.g.,$O_{T1}$)
& EWPD ($T$) \cite{Baak:2014ora,deBlas:2019qco}
& $|c_T|\!\lesssim\!10^{-3}$
& $|c_T|\!\lesssim\!\mathcal{O}(1)$
(Fig.\ref{fig:T_operator_bounds})
& \textbf{EWPD} \\

$\varphi^4 D^2$ ($hVV$ coupling)
& Higgs rates \cite{ATLAS:2019nkf,CMS:2021klu}
& $|\delta\kappa_V|\!\lesssim\!0.05$--$0.1$
& $|c_i|\!\lesssim\!\mathcal{O}(0.1$--$1)$
(alignment dependent)
& \textbf{Higgs} \\

$\varphi^4 D^2$ (aQGC in VBF)
& VBF $VVjj$ \cite{ATLAS:2022zbu,CMS:2021jji}
& $|c_i|\!\lesssim\!\mathcal{O}(10)$
& $|c_i|\!\lesssim\!\mathcal{O}(1$--$10)$
(Fig.\ref{fig:unitarity_plane})
& \textbf{Comp.} \\

$\varphi^6$ (Scalar potential)
& Di-Higgs
& Weak
& Loose for $ 
\frac{v^2}{f^2}\!\ll\!1$
(Sec.\ref{sec:scalar})
& \textbf{Unitarity} \\
\hline\hline
\end{tabular*}
\end{table}

Our analysis demonstrates that while electroweak precision and Higgs coupling measurements typically provide the most stringent bounds on dimension-6 operators accessible at low energies, unitarity constraints are essential for ensuring the self-consistency of the effective field theory at high scales. For operators that are poorly constrained by current experiments—particularly those in the $\varphi^6$ class—unitarity provides leading, model-independent limits on the viable parameter space of the 2HDMEFT.

\newpage


\section{Summary}

In this paper, we have systematically investigated the implications of perturbative unitarity on the dimension-six bosonic operators of the Two-Higgs-Doublet Model Effective Field Theory (2HDMEFT). 
By computing a comprehensive set of $2\to 2$ scattering amplitudes---including those involving longitudinally polarized gauge bosons---that exhibit growth with the center-of-mass energy, we derived rigorous constraints on the Wilson coefficients and the new-physics scale $f$.
After applying partial-wave unitarity bounds to the coupled-channel scattering matrix, we determined the energy regime in which the effective description remains self-consistent, thereby establishing a high validity scale for the model.

Our analysis reveals that unitarity constraints are especially effective in regions of parameter space where low-energy experimental probes are less sensitive. 
We demonstrate that combining unitarity limits with precision electroweak measurements---specifically the $T$-parameter---produces stringent, complementary bounds on the Wilson coefficients of custodial-symmetry-violating operators.
This synergy enables us to partially exclude certain correlated combinations of coefficients that would otherwise remain unconstrained, refining the viable parameter space of 2HDMEFT and illustrating the critical role of high-energy consistency in shaping the phenomenology of extended Higgs sectors.

\section{Appendix}
\subsection{The Potential}
\label{app:potential}
The  total potential is given as: 
$ V(\varphi_1,\varphi_2) + \mathcal{L}_{\varphi^6}$ Where, $V(\varphi_1,\phi_2)$ is given in eqn.(2.2.4)  and
\begin{eqnarray*}
\mathcal{L}_{\varphi^6} &=& \frac{1}{f^2}\Big[c_{111} |\varphi_1|^6 + c_{222} |\varphi_2|^6 
+ c_{112} |\varphi_1|^4 |\varphi_2|^2 + c_{122} |\varphi_1|^2 |\varphi_2|^4 \\
&&+ c_{(1221)1} |\varphi_1^{\dagger} \varphi_2|^2 |\varphi_1|^2 + c_{(1221)2} |\varphi_1^{\dagger} \varphi_2|^2 |\varphi_2|^2 \\
&&+ c_{(1212)1} ((\varphi_1^{\dagger} \varphi_2)^2 + h.c.) |\varphi_1|^2 
+ c_{(1212)2}  ((\varphi_1^{\dagger} \varphi_2)^2 + h.c.) |\varphi_2|^2 \\
&&+ \textcolor{blue}{c_{(1221)12} |\varphi_1^{\dagger} \varphi_2|^2 (\varphi_1^{\dagger} \varphi_2 + h.c.)} 
+ \textcolor{blue}{c_{11(12)} |\varphi_1|^4 (\varphi_1^{\dagger} \varphi_2 + h.c.)} \\
&&+ \textcolor{blue}{c_{22(12)} |\varphi_2|^4 (\varphi_1^{\dagger} \varphi_2 + h.c.)} 
+ \textcolor{blue}{c_{12(12)} |\varphi_1|^2 |\varphi_2|^2 (\varphi_1^{\dagger} \varphi_2 + h.c.)}\\ 
&&+ \textcolor{blue}{c_{121212} (\varphi_1^{\dagger} \varphi_2 + h.c.)^3}  \Big].
\end{eqnarray*}
Where, we have marked the $Z_2$-violating operators in blue colour. The minimisation conditions of this potential are: 
\begin{eqnarray*}
&&\frac{3}{4}v_1^4 c_{111} + \frac{v_1^2 v_2^2}{2}c_{112} + \frac{v_2^4}{4}c_{122}+ v_1^2 v_2^2 c_{(1212)1} + \frac{v_2^4}{2}c_{(1212)2}+ \frac{v_1^2 v_2^2}{2}c_{(1221)1}  + \frac{v_2^4}{4}c_{(1221)2} \nn\\
&&+ \frac{3}{4}v_1 v_2^3 c_{(1221)12} + \frac{5}{4}v_1^3 v_2 c_{11(12)} + \frac{3}{4}v_1 v_2^3 c_{12(12)} + \frac{v_2^5}{4 v_1}c_{22(12)} + 3 v_1 v_2^3 c_{121212}  = 0 \, ,\\
\text{and}\\
&&\frac{3}{4}v_{2}^4 c_{222} + \frac{v_1^4}{4}c_{112} + \frac{v_1^2 v_2^2}{2}c_{122} +\frac{v_1^4}{2}c_{(1212)1} + v_1^2 v_2^2 c_{(1212)2}+ \frac{v_1^4}{4} c_{(1221)1} + \frac{v_1^2 v_2^2}{2}c_{(1221)2} \nn\\
&&+\frac{3}{4}v_1^3 v_2 c_{(1221)12} + \frac{5}{4} v_1 v_2^3 c_{22(12)} + \frac{3}{4} v_1^3 v_2 c_{(12)12} + 3 v_1^3 v_2 c_{121212} = 0\, .
\end{eqnarray*}
\newpage
\subsection{Rescaling of the kinetic terms}
\label{app:phi4D2}
These operators lead to the rescaling of the kinetic terms of all the Higgs fields, without
the charged scalars. Such effects should be taken care of by appropriate field redefinitions,
which lead to the scaling of the couplings of the SM-like Higgs.
\begin{eqnarray*}
\mathcal{L}_{\varphi^4 D^2} &=& \frac{1}{f^2}\big[C_{H1} O_{H1} + C_{H2} O_{H2}
+ C_{H12} O_{H12} + C_{H1H2} O_{H1H2}\\
&& +\textcolor{blue} {C_{H1H12}} \textcolor{blue}{O_{H1H12}} + \textcolor{blue}{C_{H2H12}} \textcolor{blue}{O_{H2H12}}+  C_{T1} O_{T1}+ C_{T2} O_{T2}+ C_{T3} O_{T3}+\textcolor{blue} {C_{T4}} \textcolor{blue}{O_{T4}}+ \textcolor{blue}{C_{T5}} \textcolor{blue}{O_{T5}}\big],
\end{eqnarray*}
Where,
\begin{eqnarray*}
\label{ops}
O_{H1} &=& (\partial_{\m}|\varphi_1|^2)^2, \hspace{10pt} O_{H2} = (\partial_{\m}|\varphi_2|^2)^2,\hspace{10pt} O_{H12} = (\partial_{\m}(\varphi_1^{\dagger} \varphi_2 + h.c.))^2, \\
O_{H1H2} &=& \partial_{\m}|\varphi_1|^2 \partial^{\m}|\varphi_2|^2, O_{H1H12} = \partial_{\m}|\varphi_1|^2\partial^{\m}(\varphi_1^{\dagger} \varphi_2 + h.c.), O_{H2H12} = \partial_{\m}|\varphi_2|^2\partial^{\m}(\varphi_1^{\dagger} \varphi_2 + h.c.).
\end{eqnarray*}
Operators $O_{H1H12}$ and $O_{H2H12}$ are odd under the $Z_2$-symmetry, whereas the rest are even.

Here, we neglect the contribution from  $\varphi^2 X^2$ and $\varphi^2 D^2 X^2$ types operator becuase Constraints from electroweak precision test(EWPT) for these operators insignificant  for our purpose.

 \subsection{Scattering amplitudes}
 \label{amplitudes}
Here we derive the matrices given by the zeroth modes of partial wave amplitudes for various VBS processes for dimension-six operator in 2HDMs.
 \subsubsection{Neutral two-body states}
 For the matrix for partial wave amplitudes of neutral two-body states. Initial and final states are given by fourteen states, namely,
\begin{eqnarray*}
1=|w_1^+w_1^->,~
2=|w_2^+w_2^->,~3=|\frac{z_1z_1}{\sqrt{2}}>,~4=|\frac{z_2z_2}{\sqrt{2}}>,
~5=|\frac{h_1h_1}{\sqrt{2}}>,
\nn
\end{eqnarray*}
\begin{eqnarray*}
~6=|\frac{h_2h_2}{\sqrt{2}}>,~7=|w_1^+w_2^->,~8=|w_2^+w_1^->,~9=|h_1z_2>,~10=|h_2z_1>,\nn
\end{eqnarray*}
\begin{eqnarray*}
~11=|z_1z_2>,~12=|h_1h_2>,~13=|h_1z_1>,~14=|h_2z_2>.
\end{eqnarray*}

The elements of neutral sector $14\times 14$ matrix are given by
\begin{eqnarray*}
\mathcal{M}^{N}_{1,1}&=&4 \left(\frac{3 {c_{111}} {v_1}^2}{2 f^2}+\frac{{c_{112}} {v_2}^2}{2 f^2}\right)+\frac{2 u {C_{H1}} }{f^2},\nn\\
\mathcal{M}^{N}_{1,2}&=&\frac{{c_{112}} {v_1}^2}{f^2}+\frac{{c_{122}} {v_2}^2}{f^2}+\frac{{c_{(1221)1}} {v_1}^2}{2 f^2}+\frac{{c_{(1212)2}} {v_2}^2}{2 f^2}+\frac{2 s {C_{H12}} }{f^2}+\frac{ s {C_{H1H2}} }{2 f^2},\nn\\
\mathcal{M}^{N}_{1,3}&=& \sqrt{2} \left(\frac{3 {c_{111}} {v_1}^2}{2 f^2}+\frac{{c_{112}} {v_2}^2}{2 f^2}+\frac{{c_{(1221)1}} {v_2}^2}{4 f^2}-\frac{{c_{1212)1}} {v_2}^2}{2 f^2}\right)+\frac{2 s{C_{H1}} }{f^2},\nn\\
\mathcal{M}^{N}_{1,4}&=&
\sqrt{2} \left(\frac{9 {c_{111}} {v_1}^2}{2 f^2}+\frac{{c_{112}} {v_2}^2}{2 f^2}+\frac{{c_{(1221)1}} {v_2}^2}{4 f^2}+\frac{{c_{(1212)1}}
{v_2}^2}{2 f^2}\right)+\frac{2 s C_{H1}} {f^2},\nn\\
\mathcal{M}^{N}_{1,5}&=&
\sqrt{2} \left(\frac{c_{112} {v_1}^2}{2 f^2}+\frac{c_{122} {v_2}^2}{2 f^2}+\frac{c_{(1221)1} {v_1}^2}{4 f^2}-\frac{c_{(1212)1} {v_1}^2}{2 f^2}\right)+\frac{ s C_{H1H2} }{f^2},\nn\\
\mathcal{M}^{N}_{1,6}&=&
\sqrt{2} \left(\frac{c_{112} {v_1}^2}{2 f^2}+\frac{3 c_{122} {v_2}^2}{2 f^2}+\frac{c_{(1221)1} {v_1}^2}{4 f^2}+\frac{c_{(1212)1} {v_1}^2}{2 f^2}\right)+\frac{ s C_{H1H2} }{2 f^2},\nn\\
\mathcal{M}^{N}_{1,7}&=&
\frac{c_{(1221)1} v_1 v_2}{2 f^2}+\frac{c_{(1212)1} v_1 v_2}{f^2},\nn\\
\mathcal{M}^{N}_{1,8}&=&
\frac{c_{(1221)1} v_1 v_2}{2 f^2}+\frac{c_{(1212)1} v_1 v_2}{f^2},\nn\\
\mathcal{M}^{N}_{1,9}&=&\mathcal{M}^{N}_{1,10}=0,\nn\\
\mathcal{M}^{N}_{1,11}&=&\frac{2 c_{(1212)1} v_1 v_2}{f^2},\nn\\
\mathcal{M}_{1,12}&=&\frac{2 c_{112} v_1 v_2}{f^2}+\frac{c_{(1221)1} v_1 v_2}{f^2}+\frac{2 c_{(1212)1} v_1 v_2}{f^2},\nn\\
\mathcal{M}^{N}_{1,13}&=&\mathcal{M}^{N}_{1,14}=0,\nn\\
\mathcal{M}^{N}_{2,2}&=&  \frac{2 C_{H_2} u}{f^2}+ 4 \left(\frac{c_{122} v_1^2}{2 f^2}+\frac{3 c_{222} v_2^2}{2 f^2}\right),\nn\\
\mathcal{M}^{N}_{2,3}&=& \frac{2 C_{H1H2} s}{f^2}+\sqrt{2} \left(\frac{c_{112} v_1^2}{2 f^2}+\frac{c_{122} v_2^2}{2 f^2}+\frac{c_{(1221)2} v_2^2}{4 f^2}-\frac{c_{(1212)2} v_2^2}{2 f^2}\right)\nn\\
\mathcal{M}^{N}_{2,4}&=& \frac{ C_{H_1H_2} s}{2 f^2}+\sqrt{2} \left(\frac{3 c_{112} v_1^2}{2 f^2}+\frac{c_{122} v_2^2}{2 f^2}+\frac{c_{(1221)2} v_2^2}{4 f^2}+\frac{c_{(1212)2} v_2^2}{2 f^2}+\right)\nn\\
\mathcal{M}^{N}_{2,5}&=&\frac{2 C_{H_2} s}{f^2}+\sqrt{2} \left(\frac{c_{122} v_1^2}{2 f^2}+\frac{c_{(1221)2} v_1^2}{4 f^2}-\frac{c_{(1212)2} v_1^2}{2 f^2}+\frac{3 c_{222} v_2^2}{2 f^2}\right)\nn\\
\mathcal{M}^{N}_{2,6}&=&\frac{2 C_{H_2} s}{f^2}+\sqrt{2} \left(\frac{c_{122} v_1^2}{2 f^2}+\frac{c_{(1221)2} v_1^2}{4 f^2}+\frac{c_{(1212)2} v_1^2}{2 f^2}+\frac{9 c_{222} v_2^2}{2 f^2}\right)\nn\\
\mathcal{M}^{N}_{2,7}&=& \left(\frac{c_{(1221)2} v_1 v_2}{2 f^2}+\frac{c_{(1212)2} v_1 v_2}{f^2}\right),\nn\\
\mathcal{M}^{N}_{2,8}&=& \left(\frac{c_{(1221)2} v_1 v_2}{2 f^2}+\frac{c_{(1212)2} v_1 v_2}{f^2}\right)\nn\\
\mathcal{M}^{N}_{2,9}&=&\mathcal{M}^{N}_{2,10}=0,\nn\\
\mathcal{M}^{N}_{2,11}&=&\frac{2 c_{(1212)2} v_1 v_2}{f^2},\nn\\
\end{eqnarray*}
\begin{eqnarray*}
\mathcal{M}^{N}_{2,12}&=&\frac{2 c_{122} v_1 v_2}{f^2}+\frac{c_{(1221)2} v_1 v_2}{f^2}+\frac{2 c_{(1212)2} v_1 v_2}{f^2},\nn\\
\mathcal{M}^{N}_{2,13}&=&\mathcal{M}^{N}_{2,14}=0,\nn\\
\mathcal{M}^{N}_{3,3}&=& 12 \left(\frac{3 c_{111} v_1^2}{8 f^2}+\frac{c_{112} v_2^2}{8 f^2}+\frac{c_{(1221)1} v_2^2}{8 f^2}-\frac{c_{(1212)1} v_2^2}{4 f^2}+\frac{\lambda _1}{4}+\frac{\lambda _3}{4}\right),\nn\\
\mathcal{M}^{N}_{3,4}&=&\frac{2 C_{H_1} s}{f^2}+\frac{36 c_{111} v_1^2+4 c_{112} v_2^2+4 c_{(1221)1} v_2^2+8 f^2 \lambda _1+8 f^2 \lambda _3}{8 f^2},\nn\\
\mathcal{M}^{N}_{3,5}&=& \frac{C_{H_1H_2} s}{f^2}+\frac{2 c_{112} v_1^2+2 c_{(1221)1} v_1^2+2 c_{122} v_2^2+2 c_{(1221)2} v_2^2+4 f^2 \lambda _3+2 f^2 \lambda _5}{4 f^2},\nn\\
\mathcal{M}^{N}_{3,6}&=& \frac{C_{H_1H_2} s}{f^2}+\frac{2 c_{112} v_1^2+2 c_{(1221)1} v_1^2+6 c_{122} v_2^2+6 c_{(1221)2} v_2^2-12 c_{(1212)2} v_2^2+4 f^2 \lambda _3+2 f^2 \lambda _6}{4 f^2},\nn\\
\mathcal{M}^{N}_{3,7}&=&\mathcal{M}^{N}_{3,8}=\mathcal{M}^{N}_{3,9}=\mathcal{M}^{N}_{3,10}=\mathcal{M}^{N}_{3,11}=\mathcal{M}^{N}_{3,12}=\mathcal{M}^{N}_{3,13}=\mathcal{M}^{N}_{3,14}=0,\nn\\
\mathcal{M}^{N}_{4,4}&=& 12 \left(\frac{c_{112} v_2^2}{8 f^2}+\frac{c_{(1221)1} v_2^2}{8 f^2}+\frac{c_{(1212)1} v_2^2}{4 f^2}+\frac{15 \text{C$\_$} v_1^2}{8 f^2}+\frac{\lambda _1}{4}+\frac{\lambda _3}{4}\right),\nn\\
\mathcal{M}^{N}_{4,5}&=&\frac{C_{H_1H_2} s}{f^2}+\frac{4 c_{112} v_2^2+4 c_{(1221)1} v_2^2+36 \text{C$\_$} v_1^2+8 f^2 \lambda _1+8 f^2 \lambda _3}{8 f^2},\nn\\
\mathcal{M}^{N}_{4,6}&=&\frac{C_{H_1H_2} s}{f^2}+\frac{2 c_{112} v_1^2+2 c_{(1221)1} v_1^2+2 c_{122} v_2^2+2 c_{(1212)1} v_2^2+4 f^2 \lambda _3+2 f^2 \lambda _5}{4 f^2},\nn\\
\mathcal{M}^{N}_{4,7}&=&\mathcal{M}^{N}_{4,8}=\mathcal{M}^{N}_{4,9}=\mathcal{M}^{N}_{4,10}=\mathcal{M}^{N}_{4,11}=\mathcal{M}^{N}_{4,12}=\mathcal{M}^{N}_{4,13}=\mathcal{M}^{N}_{4,14}=0,\nn\\
\mathcal{M}^{N}_{5,5}&=&12 \left(\frac{c_{122} v_1^2}{8 f^2}+\frac{c_{(1212)1} v_1^2}{8 f^2}-\frac{c_{(1212)2} v_1^2}{4 f^2}+\frac{3 c_{222} v_2^2}{8 f^2}+\frac{\lambda _2}{4}+\frac{\lambda _3}{4}\right),\nn\\
\mathcal{M}^{N}_{5,6}&=&\frac{2 C_{H_2} s}{f^2}+\frac{4 c_{122} v_1^2+4 c_{(1212)1} v_1^2+36 c_{222} v_2^2+8 f^2 \lambda _2+8 f^2 \lambda _3}{8 f^2},\nn\\
\mathcal{M}^{N}_{5,7}&=&\mathcal{M}^{N}_{5,8}=\mathcal{M}^{N}_{5,9}=\mathcal{M}^{N}_{5,10}=\mathcal{M}^{N}_{5,11}=\mathcal{M}^{N}_{5,12}=\mathcal{M}^{N}_{5,13}=\mathcal{M}^{N}_{5,14}=0,\nn\\
\mathcal{M}^{N}_{6,6}&=&12 \left(\frac{c_{122} v_1^2}{8 f^2}+\frac{c_{(1212)1} v_1^2}{8 f^2}+\frac{c_{(1212)2} v_1^2}{4 f^2}+\frac{15 c_{222} v_2^2}{8 f^2}+\frac{\lambda _2}{4}+\frac{\lambda _3}{4}\right),\nn\\
\mathcal{M}^{N}_{6,7}&=&\mathcal{M}^{N}_{6,8}=\mathcal{M}^{N}_{6,9}=\mathcal{M}^{N}_{6,10}=\mathcal{M}^{N}_{6,11}=\mathcal{M}^{N}_{6,12}=\mathcal{M}^{N}_{6,13}=\mathcal{M}^{N}_{6,14}=0,\nn\\
\mathcal{M}^{N}_{7,7}&=& \frac{C_{H_1H_2} t}{f^2}+4 \left(\frac{c_{112} v_1^2}{f^2}+\frac{c_{(1221)1} v_1^2}{2 f^2}+\frac{c_{122} v_2^2}{f^2}+\frac{c_{(1212)1} v_2^2}{2 f^2}+2 \lambda _3+\frac{\lambda _5}{2}+\frac{\lambda _6}{2}\right),\nn\\
\mathcal{M}^{N}_{7,8}&=&\frac{C_{H_{12}} t}{f^2}+\frac{c_{(1212)1} v_1^2}{2 f^2}+\frac{c_{(1212)2} v_2^2}{2 f^2}+\frac{\lambda _5}{4}-\frac{\lambda _6}{4},\nn\\
\mathcal{M}^{N}_{7,9}&=& \frac{3 i c_{(1221)1} v_1^2}{4 f^2}-\frac{3 i c_{(1212)1} v_1^2}{2 f^2}+\frac{i c_{(1212)1} v_2^2}{4 f^2}-\frac{i c_{(1212)2} v_2^2}{2 f^2}+\frac{i \lambda _6}{2}-\frac{i \lambda _4}{2},\nn\\
\mathcal{M}^{N}_{7,10}&=&\frac{i c_{(1212)1} v_1^2}{2 f^2}-\frac{i c_{(1221)1} v_1^2}{4 f^2}+\frac{3 i c_{(1212)2} v_2^2}{2 f^2}-\frac{3 i c_{(1212)1} v_2^2}{4 f^2}+\frac{i \lambda _4}{2}-\frac{i \lambda _6}{2},\nn\\
\mathcal{M}^{N}_{7,11}&=&\frac{2 C_{H_{12}} s}{f^2}+\frac{c_{(1221)1} v_1^2}{4 f^2}+\frac{c_{(1212)1} v_1^2}{2 f^2}+\frac{c_{(1212)1} v_2^2}{4 f^2}+\frac{c_{(1212)2} v_2^2}{2 f^2}+\frac{\lambda _5}{2}-\frac{\lambda _4}{2},\nn\\
\end{eqnarray*}
\begin{eqnarray*}
\mathcal{M}^{N}_{7,12}&=&\frac{2 C_{H_{12}} s}{f^2}+\frac{3 c_{(1221)1} v_1^2}{4 f^2}+\frac{3 c_{(1212)1} v_1^2}{2 f^2}+\frac{3 c_{(1212)1} v_2^2}{4 f^2}+\frac{3 c_{(1212)2} v_2^2}{2 f^2}+\frac{\lambda _5}{2}-\frac{\lambda _4}{2},\nn\\
\mathcal{M}^{N}_{7,13}&=&\frac{i c_{(1221)1} v_1 v_2}{2 f^2}-\frac{i c_{(1212)1} v_1 v_2}{f^2},\nn\\
\mathcal{M}^{N}_{7,14}&=&\frac{i c_{(1212)2} v_1 v_2}{f^2}-\frac{i c_{(1212)1} v_1 v_2}{2 f^2},\nn\\
\mathcal{M}^{N}_{8,8}&=&\frac{C_{H1H2} t}{f^2}+4 \left(\frac{c_{112} v_1^2}{f^2}+\frac{c_{(1221)1} v_1^2}{2 f^2}+\frac{c_{122} v_2^2}{f^2}+\frac{c_{(1212)1} v_2^2}{2 f^2}+2 \lambda _3+\frac{\lambda _5}{2}+\frac{\lambda _6}{2}\right),\nn\\
\mathcal{M}^{N}_{8,9}&=&\frac{3 i c_{(1212)1} v_1^2}{2 f^2}-\frac{3 i c_{(1221)1} v_1^2}{4 f^2}+\frac{i c_{(1212)2} v_2^2}{2 f^2}-\frac{i c_{(1212)1} v_2^2}{4 f^2}+\frac{i \lambda _4}{2}-\frac{i \lambda _6}{2},\nn\\
\mathcal{M}^{N}_{8,10}&=&\frac{i c_{(1221)1} v_1^2}{4 f^2}-\frac{i c_{(1212)1} v_1^2}{2 f^2}+\frac{3 i c_{(1212)1} v_2^2}{4 f^2}-\frac{3 i c_{(1212)2} v_2^2}{2 f^2}+\frac{i \lambda _6}{2}-\frac{i \lambda _4}{2},\nn\\
\mathcal{M}^{N}_{8,11}&=&\frac{2 C_{H12} s}{f^2}+\frac{i c_{(1221)1} v_1^2}{4 f^2}-\frac{i c_{(1212)1} v_1^2}{2 f^2}+\frac{3 i c_{(1212)1} v_2^2}{4 f^2}-\frac{3 i c_{(1212)2} v_2^2}{2 f^2}+\frac{i \lambda _6}{2}-\frac{i \lambda _4}{2},\nn\\
\mathcal{M}^{N}_{8,12}&=&\frac{2 C_{H12} s}{f^2}+ \frac{c_{(1221)1} v_1^2}{4 f^2}+\frac{c_{(1212)1} v_1^2}{2 f^2}+\frac{c_{(1212)1} v_2^2}{4 f^2}+\frac{c_{(1212)2} v_2^2}{2 f^2}+\frac{\lambda _5}{2}-\frac{\lambda _4}{2},\nn\\
\mathcal{M}^{N}_{8,13}&=&\frac{i c_{(1212)1} v_1 v_2}{f^2}-\frac{i c_{(1221)1} v_1 v_2}{2 f^2},\nn\\
\mathcal{M}^{N}_{8,14}&=&\frac{i c_{(1212)1} v_1 v_2}{2 f^2}-\frac{i c_{(1212)2} v_1 v_2}{f^2},\nn\\
\mathcal{M}^{N}_{9,9}&=& \frac{C_{H1H2} t}{f^2} +4 \left(\frac{3 c_{112} v_1^2}{4 f^2}+\frac{3 c_{(1221)1} v_1^2}{4 f^2}-\frac{3 c_{(1212)1} v_1^2}{2 f^2}+\frac{c_{122} v_2^2}{4 f^2}+\frac{c_{(1212)1} v_2^2}{4 f^2}+\frac{\lambda _3}{2}+\frac{\lambda _6}{4}\right),\nn\\
\mathcal{M}^{N}_{9,10}&=&\frac{2 C_{H12} t}{f^2}+\frac{3 c_{(1212)1} v_1^2}{f^2}+\frac{3 c_{(1212)2} v_2^2}{f^2}+\frac{\lambda _5}{2}-\frac{\lambda _6}{2},\nn\\
\mathcal{M}^{N}_{9,11}&=&\mathcal{M}^{N}_{9,12}=0,\nn\\
\mathcal{M}^{N}_{9,13}&=&
\frac{6 c_{(1212)1} v_1 v_2}{f^2},\nn\\
\mathcal{M}_{9,14}&=&
2 \left(\frac{c_{122} v_1 v_2}{f^2}+\frac{c_{(1212)1} v_1 v_2}{f^2}\right),\nn\\
\mathcal{M}^{N}_{10,10}&=&\frac{2 C_{H12} t}{f^2}+4 \left(\frac{c_{112} v_1^2}{4 f^2}+\frac{c_{(1221)1} v_1^2}{4 f^2}+\frac{3 c_{122} v_2^2}{4 f^2}+\frac{3 c_{(1212)1} v_2^2}{4 f^2}-\frac{3 c_{(1212)2} v_2^2}{2 f^2}+\frac{\lambda _3}{2}+\frac{\lambda _6}{4}\right),\nn\\
\mathcal{M}^{N}_{10,11}&=&\mathcal{M}^{N}_{10,12}=0,\nn\\
\mathcal{M}^{N}_{10,13}&=&2 \left(\frac{c_{112} v_1 v_2}{f^2}+\frac{c_{(1221)1} v_1 v_2}{f^2}\right),\nn\\
\mathcal{M}^{N}_{10,14}&=&\frac{3 c_{(1212)2} v_1 v_2}{f^2},\nn\\
\end{eqnarray*}
\begin{eqnarray*}
\mathcal{M}^{N}_{11,11}&=&\frac{C_{H12} t}{f^2}+\frac{C_{H1H2} t}{f^2}+4 \left(\frac{c_{112} v_1^2}{4 f^2}+\frac{c_{(1221)1} v_1^2}{4 f^2}+\frac{c_{122} v_2^2}{4 f^2}+\frac{c_{(1212)1} v_2^2}{4 f^2}+\frac{\lambda _3}{2}+\frac{\lambda _5}{4}\right),\nn\\
\mathcal{M}^{N}_{11,12}&=&\frac{2 C_{H12} s}{f^2}+\frac{3 c_{(1212)1} v_1^2}{f^2}+\frac{3 c_{(1212)2} v_2^2}{f^2}+\frac{\lambda _5}{2}-\frac{\lambda _6}{2},\nn\\
\mathcal{M}^{N}_{11,13}&=&
\mathcal{M}^{N}_{11,14}=0,\nn\\
\mathcal{M}^{N}_{12,12}&=&\frac{C_{H12} t}{f^2}+\frac{C_{H1H2} t}{f^2}+4 \left(\frac{3 c_{112} v_1^2}{4 f^2}+\frac{3 c_{(1221)1} v_1^2}{4 f^2}+\frac{3 c_{(1212)1} v_1^2}{2 f^2}\right)\nn\\
&&+4 \left(\frac{3 c_{122} v_2^2}{4 f^2}+\frac{3 c_{(1212)1} v_2^2}{4 f^2}+\frac{3 c_{(1212)2} v_2^2}{2 f^2}+\frac{\lambda _3}{2}+\frac{\lambda _5}{4}\right),\nn\\
\mathcal{M}^{N}_{12,13}&=&
\mathcal{M}^{N}_{12,14}=0,\nn\\
\mathcal{M}^{N}_{13,13}&=&\frac{2 C_{H1} t}{f^2}+4 \left(\frac{c_{112} v_2^2}{4 f^2}+\frac{c_{(1221)1} v_2^2}{4 f^2}+\frac{9 c_{111}v_1^2}{4 f^2}+\frac{\lambda _1}{2}+\frac{\lambda _3}{2}\right),\nn\\
\mathcal{M}^{N}_{13,14}&=&\frac{2 C_{H12} t}{f^2}+\frac{3 c_{(1212)1} v_1^2}{f^2}+\frac{3 c_{(1212)2} v_2^2}{f^2}+\frac{\lambda _5}{2}-\frac{\lambda _6}{2},\nn\\
\mathcal{M}^{N}_{14,14}&=&\frac{2 C_{H2} t}{f^2}+4 \left(\frac{c_{122} v_1^2}{4 f^2}+\frac{c_{(1212)1} v_1^2}{4 f^2}+\frac{9 c_{222} v_2^2}{4 f^2}+\frac{\lambda _2}{2}+\frac{\lambda _3}{2}\right).\nn
\end{eqnarray*}

\subsubsection{Charged two-body states}
For the singly charged sector, the S-matrix is $8\times8$ matrix with the following two-particle states as rows and columns:
\begin{eqnarray*}
1=|h_1 w_1^+>,~
2=|h_1 w_2^+>,~3=|z_1 w_1^+>,
~4=|z_2 w_2^+>,\\
5=|h_1 w_2^+>,~
6=|h_2 w_1^+>,~7=|z_1 w_2^+>,
~8=|z_2 w_1^+>.\nn\\
\end{eqnarray*}

The elements of the matrix are given by:
\begin{eqnarray*}
\mathcal{M}^{C}_{1,1}&=& \frac{2 C_{H2} t}{f^2}+2 \left(\frac{9 c_{111} v_1^2}{2 f^2}+\frac{c_{112} v_2^2}{2 f^2}+\frac{c_{(1221)1} v_2^2}{4 f^2}+\frac{c_{(1212)1} v_2^2}{2 f^2}+\lambda _1+\lambda _3\right),\nn\\
\mathcal{M}^{C}_{1,2}&=& \frac{2 C_{H12} t}{f^2}+2 \left(\frac{9 c_{111} v_1^2}{2 f^2}+\frac{c_{112} v_2^2}{2 f^2}+\frac{c_{(1221)1} v_2^2}{4 f^2}+\frac{c_{(1212)1} v_2^2}{2 f^2}+\lambda _1+\lambda _3\right),\nn\\
\mathcal{M}^{C}_{1,3}&=&0,\nn\\
\mathcal{M}^{C}_{1,4}&=&\frac{3 i c_{(1221)1} v_1^2}{4 f^2}-\frac{3 i c_{(1212)1} v_1^2}{2 f^2}+\frac{i c_{(1212)1} v_2^2}{4 f^2}-\frac{i c_{(1212)2} v_2^2}{2 f^2}+\frac{i \lambda _6}{2}-\frac{i \lambda _4}{2},\nn\\
\mathcal{M}^{C}_{1,5}&=&\frac{3 c_{(1221)1} v_1 v_2}{4 f^2}+\frac{3 c_{(1212)1} v_1 v_2}{2 f^2},\nn\\
\end{eqnarray*}
\begin{eqnarray*}
\mathcal{M}^{C}_{1,6}&=&
\frac{3 c_{(1221)1} v_1^2}{4 f^2}+\frac{3 c_{(1212)1} v_1^2}{2 f^2}+\frac{3 c_{(1212)1} v_2^2}{4 f^2}+\frac{3 c_{(1212)2} v_2^2}{2 f^2}+\frac{\lambda _5}{2}-\frac{\lambda _4}{2},\nn\\
\mathcal{M}^{C}_{1,7}&=&\frac{i c_{(1221)1} v_1^2}{4 f^2}-\frac{i c_{(1212)1} v_1^2}{2 f^2}+\frac{3 i c_{(1212)1} v_2^2}{4 f^2}-\frac{3 i c_{(1212)2} v_2^2}{2 f^2}+\frac{i \lambda _6}{2}-\frac{i \lambda _4}{2},\nn\\
\mathcal{M}^{C}_{1,8}&=&\frac{3 i c_{(1212)1} v_1^2}{2 f^2}-\frac{3 i c_{(1221)1} v_1^2}{4 f^2}+\frac{i c_{(1212)2} v_2^2}{2 f^2}-\frac{i c_{(1212)1} v_2^2}{4 f^2}+\frac{i \lambda _4}{2}-\frac{i \lambda _6}{2},\nn\\
\mathcal{M}^{C}_{2,2}&=&\frac{2 C_{H2} t}{f^2}2 +\left(\frac{c_{122} v_1^2}{2 f^2}+\frac{c_{(1212)1} v_1^2}{4 f^2}+\frac{c_{(1212)2} v_1^2}{2 f^2}+\frac{9 c_{222} v_2^2}{2 f^2}+\lambda _2+\lambda _3\right),\nn\\
\mathcal{M}^{C}_{2,3}&=&\frac{i c_{(1221)1} v_1^2}{4 f^2}-\frac{i c_{(1212)1} v_1^2}{2 f^2}+\frac{3 i c_{(1212)1} v_2^2}{4 f^2}-\frac{3 i c_{(1212)2} v_2^2}{2 f^2}+\frac{i \lambda _6}{2}-\frac{i \lambda _4}{2},\nn\\
\mathcal{M}^{C}_{2,4}&=&0,\nn\\
\mathcal{M}^{C}_{2,5}&=&\frac{3 c_{112} v_1^2}{2 f^2}+\frac{c_{122} v_2^2}{2 f^2}+\frac{c_{(1212)1} v_2^2}{4 f^2}+\frac{c_{(1212)2} v_2^2}{2 f^2}+\lambda _3+\frac{\lambda _4}{2},\nn\\
\mathcal{M}^{C}_{2.6}&=&\frac{3 c_{(1212)1} v_1 v_2}{4 f^2}+\frac{3 c_{(1212)2} v_1 v_2}{2 f^2},\nn\\
\mathcal{M}^{C}_{2,7}&=&0,\nn\\
\mathcal{M}^{C}_{2,8}&=&\frac{i c_{(1212)1} v_1 v_2}{2 f^2}-\frac{i c_{(1212)2} v_1 v_2}{f^2},\nn\\
\mathcal{M}^{C}_{3,3}&=&\frac{2 C_{H1} t}{f^2}+2 \left(\frac{3 c_{111} v_1^2}{2 f^2}+\frac{c_{112} v_2^2}{2 f^2}+\frac{c_{(1221)1} v_2^2}{4 f^2}-\frac{c_{(1212)1} v_2^2}{2 f^2}+\lambda _1+\lambda _3\right),\nn\\
\mathcal{M}^{C}_{3,4}&=&\frac{2 C_{H12} t}{f^2}+\frac{c_{(1221)1} v_1^2}{4 f^2}+\frac{c_{(1212)1} v_1^2}{2 f^2}+\frac{c_{(1212)1} v_2^2}{4 f^2}+\frac{c_{(1212)2} v_2^2}{2 f^2}+\frac{\lambda _5}{2}-\frac{\lambda _4}{2},\nn\\
\mathcal{M}^{C}_{3,5}&=&\frac{3 i c_{(1212)1} v_1^2}{2 f^2}-\frac{3 i c_{(1221)1} v_1^2}{4 f^2}+\frac{i c_{(1212)2} v_2^2}{2 f^2}-\frac{i c_{(1212)1} v_2^2}{4 f^2}+\frac{i \lambda _4}{2}-\frac{i \lambda _6}{2},\nn\\
\mathcal{M}^{C}_{3,6}&=&
\mathcal{M}^{C}_{3,7}=0,\nn\\
\mathcal{M}^{C}_{3,8}&=&
\frac{2 c_{(1212)1} v_1 v_2}{f^2},\nn\\
\mathcal{M}^{C}_{4,4}&=&\frac{2 C_{H1} t}{f^2}+2 \left(\frac{c_{122} v_1^2}{2 f^2}+\frac{c_{(1212)1} v_1^2}{4 f^2}-\frac{c_{(1212)2} v_1^2}{2 f^2}+\frac{3 c_{222} v_2^2}{2 f^2}+\lambda _2+\lambda _3\right),\nn\\
\mathcal{M}^{C}_{4,5}&=&
0,\nn\\
\mathcal{M}^{C}_{4,6}&=&\frac{i c_{(1212)1} v_1 v_2}{2 f^2}-\frac{i c_{(1212)2} v_1 v_2}{f^2},\nn\\
\mathcal{M}^{C}_{4,7}&=&\frac{2 c_{(1212)2} v_1 v_2}{f^2},\nn\\
\mathcal{M}^{C}_{4,8}&=&\frac{c_{(1212)1} v_1 v_2}{4 f^2}+\frac{c_{(1212)2} v_1 v_2}{2 f^2},\nn\\
\mathcal{M}^{C}_{5,5}&=& \frac{2 C_{H1H2} t}{f^2}+\frac{3 c_{112} v_1^2}{2 f^2}+\frac{c_{122} v_2^2}{2 f^2}+\frac{c_{(1212)1} v_2^2}{4 f^2}+\frac{c_{(1212)2} v_2^2}{2 f^2}+\lambda _3+\frac{\lambda _4}{2},\nn\\
\end{eqnarray*}
\begin{eqnarray*}
\mathcal{M}^{C}_{5,6}&=&\frac{2 C_{H12} t}{f^2}+\frac{3 c_{(1221)1} v_1^2}{4 f^2}+\frac{3 c_{(1212)1} v_1^2}{2 f^2}+\frac{3 c_{(1212)1} v_2^2}{4 f^2}+\frac{3 c_{(1212)2} v_2^2}{2 f^2}+\frac{\lambda _5}{2}-\frac{\lambda _4}{2},\nn\\
\mathcal{M}^{C}_{5,7}&=&0,\nn\\
\mathcal{M}^{C}_{5,8}&=&
\frac{3 i c_{(1212)1} v_1^2}{2 f^2}-\frac{3 i c_{(1221)1} v_1^2}{4 f^2}+\frac{i c_{(1212)2} v_2^2}{2 f^2}-\frac{i c_{(1212)1} v_2^2}{4 f^2}+\frac{i \lambda _4}{2}-\frac{i \lambda _6}{2},\nn\\
\mathcal{M}^{C}_{6,6}&=&\frac{2 C_{H1H2} t}{f^2}+\frac{c_{112} v_1^2}{2 f^2}+\frac{c_{(1221)1} v_1^2}{4 f^2}+\frac{c_{(1212)1} v_1^2}{2 f^2}+\frac{3 c_{122} v_2^2}{2 f^2}+\lambda _3+\frac{\lambda _4}{2},\nn\\
\mathcal{M}^{C}_{6,7}&=&\frac{i c_{(1212)1} v_1^2}{2 f^2}-\frac{i c_{(1221)1} v_1^2}{4 f^2}+\frac{3 i c_{(1212)2} v_2^2}{2 f^2}-\frac{3 i c_{(1212)1} v_2^2}{4 f^2}+\frac{i \lambda _4}{2}-\frac{i \lambda _6}{2},\nn\\
\mathcal{M}^{C}_{6,8}&=&0,\nn\\
\mathcal{M}^{C}_{7,7}&=&\frac{2 C_{H1H2} t}{f^2}+\frac{c_{112} v_1^2}{2 f^2}+\frac{c_{122} v_2^2}{2 f^2}+\frac{c_{(1212)1} v_2^2}{4 f^2}-\frac{c_{(1212)2} v_2^2}{2 f^2}+\lambda _3+\frac{\lambda _4}{2},\nn\\
\mathcal{M}^{C}_{7,8}&=&\frac{2 C_{H12} t}{f^2}+\frac{c_{(1221)1} v_1^2}{4 f^2}+\frac{c_{(1212)1} v_1^2}{2 f^2}+\frac{c_{(1212)1} v_2^2}{4 f^2}+\frac{c_{(1212)2} v_2^2}{2 f^2}+\frac{\lambda _5}{2}-\frac{\lambda _4}{2},\nn\\
\mathcal{M}^{C}_{8,8}&=&\frac{2 C_{H1H2} t}{f^2}+\frac{c_{112} v_1^2}{2 f^2}+\frac{c_{(1221)1} v_1^2}{4 f^2}-\frac{c_{(1212)1} v_1^2}{2 f^2}+\frac{c_{122} v_2^2}{2 f^2}+\lambda _3+\frac{\lambda _4}{2} .\nn\
\end{eqnarray*}




\end{document}